\documentclass[10pt]{aastex}
\usepackage{emulateapj5,apjfonts}
\usepackage{graphicx}
\slugcomment{{\sc Accepted to ApJ:} May 14, 2005}
\newcommand{\begit}{\begin{itemize}}
\newcommand{\enit}{\end{itemize}}
\newcommand{\begen}{\begin{enumerate}}
\newcommand{\enen}{\end{enumerate}}
\newcommand{\p}{\partial}    
\newcommand{\pr}{^\prime} 
\newcommand       \be           {\begin{equation}}
\newcommand       \ee           {\end{equation}}
\newcommand       \bea          {\begin{eqnarray}}
\newcommand       \eea          {\end{eqnarray}}
\newcommand      \fgas          {f_{g_{\,0.5}}}
\newcommand      \fg            {f_{g_{\,0.1}}}
\newcommand       \kms		{\,{\rm km \,\, s}^{-1}}

\newcommand{\beqa}{\begin{eqnarray}} 
\newcommand{\eeqa}{\end{eqnarray}} 
\newcommand{\teff}{T_{\rm eff}}
\newcommand{\sds}{\dot{\Sigma}_\star}
\newcommand{\rkpc}{r_{\rm kpc}}
\newcommand{\mtot}{M_{\rm tot}}
\newcommand{\stot}{\Sigma_{\rm tot}}
\newcommand{\ssb}{\sigma_{\rm SB}}
\newcommand{\mds}{\dot{M}_\star}

\newcommand{\tauv}{\tau_{\rm V}}

\newcommand{\rtauv}{R_{\tau_{\rm V}=1}}

\newcommand{\tsub}{T_{\rm sub}}
\newcommand{\rsub}{R_{\rm sub}}
\newcommand{\ksub}{\kappa_{\rm sub}}

\def\mpy{\rm \ M_\odot \ {\rm yr^{-1}}}
\def\mpykpc{\rm \ M_\odot \ {\rm yr^{-1}} \ kpc^{-2}}
\def\lkpc{\rm \ L_\odot \ kpc^{-2}}

\begin{document}

\title{Radiation Pressure Supported Starburst Disks \& AGN Fueling}

\author{Todd A.~Thompson,\altaffilmark{1,2} Eliot Quataert,\altaffilmark{3} \& Norman Murray\altaffilmark{4,5,6}}

\altaffiltext{1}{Hubble Fellow}
\altaffiltext{2}{
Astronomy Department 
\& Theoretical Astrophysics Center, 601 Campbell Hall, 
The University of California, Berkeley, CA 94720; 
thomp@astro.berkeley.edu}
\altaffiltext{3}{Astronomy Department 
\& Theoretical Astrophysics Center, 601 Campbell Hall, 
The University of California, Berkeley, CA 94720; 
eliot@astro.berkeley.edu}
\altaffiltext{4}{Canada Research Chair in Astrophysics}
\altaffiltext{5}{Visiting Miller Professor, The University of California, Berkeley}
\altaffiltext{6}{Canadian Institute for Theoretical Astrophysics, 60 St.~George Street, University of Toronto, Toronto,
ON M5S 3H8, Canada; murray@cita.utoronto.ca}

\begin{abstract}

We consider the structure of marginally Toomre-stable starburst disks
under the assumption that radiation pressure on dust grains provides
the dominant vertical support against gravity.  This assumption is
particularly appropriate when the disk is optically thick to its own
infrared radiation, as in the central regions of Ultraluminous
Infrared Galaxies (ULIRGs).  We argue that because the disk radiates
at its Eddington limit (for dust), the ``Schmidt-law'' for star formation changes
in the optically-thick limit, with the star formation rate per unit
area scaling as $\sds\propto\Sigma_g/\kappa$, where $\Sigma_g$ is the
gas surface density and $\kappa$ is the mean opacity of the disk.  Our
calculations further show that optically thick starburst disks have a
characteristic flux, star formation rate per unit area, and dust
effective temperature of $F\sim10^{13}$ L$_\odot$ kpc$^{-2}$, $\dot
\Sigma_\star \sim 10^3 \mpykpc$, and $T_{\rm eff} \sim 90$ K,
respectively.  We compare our model predictions with observations of
ULIRGs and find good agreement.

We extend our model of starburst disks from many-hundred parsec scales
to sub-parsec scales and address the problem of fueling active
galactic nuclei (AGN).  We assume that angular momentum transport
proceeds via global torques (e.g., spiral waves, winds, or a central
bar) rather than a local viscosity. We consistently account for the
radial depletion of gas due to star formation and find a strong
bifurcation between two classes of disk models: (1) solutions with a
starburst on large scales that consumes all of the gas with little or
no fueling of a central AGN and (2) models with an outer large-scale
starburst accompanied by a more compact starburst on $1-10$ pc scales
and a bright central AGN.  The luminosity of the latter models is in
many cases dominated by the AGN, although these disk solutions exhibit
a broad mid- to far-infrared peak from star formation.  We show that
the vertical thickness of the starburst disk on parsec scales can
approach $h\sim r$, perhaps accounting for the nuclear obscuration in
some Type 2 AGN. We also argue that the disk of young stars in the
Galactic Center may be the remnant of such a compact nuclear
starburst.

\end{abstract}


\section{Introduction}
\label{section:intro}

Star formation in galaxies is observed to be globally inefficient; in
a sample of local spiral galaxies and luminous starbursts, Kennicutt
(1998) showed that only a few percent of the gas is converted into
stars each dynamical time.  This inefficiency may result from
``feedback'': the energy and momentum injected into the interstellar
medium (ISM) by star formation can in turn regulate the star formation
rate in a galaxy.  Models for feedback in the ISM generally invoke
energy and momentum injection by supernovae and stellar winds (e.g.,
McKee \& Ostriker 1977; Silk 1997; Efstathiou 2000).  However, the
momentum supplied to the ISM by the radiation from massive stars is
comparable to that supplied by supernovae and stellar winds.  The UV
radiation from massive stars is absorbed and scattered by dust grains,
which reprocess the UV emission into the IR.  Because the dust grains
are hydrodynamically coupled to the gas, radiation pressure on dust
can help stabilize the gas against its own self-gravity and may
therefore be an important feedback mechanism.  

When the ISM of a galaxy is optically thick to the re-radiated IR
emission, radiative diffusion ensures that all of the momentum from
the photons produced by star formation is efficiently coupled to the
gas. We show that this limit is applicable on scales of hundreds of
parsecs in luminous gas-rich starbursts, including Ultraluminous
Infrared Galaxies (ULIRGs), the most luminous and dust enshrouded
starbursts known (e.g., Genzel \& Cesarsky 2000).  Indeed, Scoville
(2003) has already pointed out that radiation pressure on dust could
plausibly be the dominant source of support against gravity in ULIRGs.
We quantify this hypothesis by developing models of radiation pressure
supported starburst disks.

Although radiation-pressure supported disks have not been extensively
considered in the galactic context, they have been well studied in
models of black hole accretion.  Most notably for our present
purposes, the outer parts of disks around active galactic nuclei (AGN)
are expected to be dominated by radiation pressure on dust (e.g.,
Sirko \& Goodman 2003).  In the context of this paper, the distinction
between the galactic disk and the AGN disk becomes somewhat unclear:
if luminous AGN are fueled by gas from the cold ISM of their host
galaxy, there must be a continuous transition from the star-forming
``galactic'' disk to the central black hole's ``accretion disk.''  The
nature of this transition, and indeed whether it occurs at all,
remains uncertain.  The problem is that the outer parts of AGN disks
are strongly self-gravitating with a Toomre stability parameter $Q \ll
1$ (Kolykhalov \& Sunyaev 1980; 
Shlosman \& Begelman 1989; Shlosman, Begelman, \& Frank 1990;
Kumar 1999; Goodman 2003; Levin 2005; Tan \& Blackman 2005).  It is difficult to see how the disk avoids
fragmenting almost entirely into stars.  One possibility is that in
the dense gas-rich nuclear regions of galaxies, angular momentum
transport proceeds via global torques (e.g., bars, spiral waves,
large-scale magnetic stresses, etc.), rather than a local viscosity
(Shlosman, Frank, \& Begelman 1990; Mihos \& Hernquist 1996; Goodman 2003).  In this case, gas
may inflow sufficiently rapidly to avoid turning entirely into stars.
In order to explore the fueling of a central AGN, we extend our
galactic-scale disk models to smaller radii using a parameterization
of the transport of angular momentum by global torques.  This provides
a framework for understanding the relationship between AGN activity
and nuclear starbursts.

The plan for this paper is as follows.  In
\S\ref{section:onezone}, we develop a simple model of self-regulated
star formation in galactic disks.  We distinguish between disks that
are optically thick/thin to their own IR radiation and we present
models appropriate to each of these limits.  
Throughout \S\ref{section:onezone}, and in particular in
\S\ref{section:comparisonzero}, we compare our theoretical models with
observations of starburst galaxies.  In \S\ref{section:disk}, we
extend the results of \S\ref{section:onezone} to include angular
momentum transport and we address the problem of AGN fueling.
Finally, \S\ref{section:discussion} provides a discussion and summary of
our conclusions, and highlights some of the predictions of our models.


\section{A One-Zone Disk Model}
\label{section:onezone}

In this section, we construct simple dynamical models for the
structure of star-forming disk galaxies.  We consider two limits: (1)
the ``optically-thin'' limit, in which the disk is optically-thick to
the UV radiation produced by massive stars, but optically-thin to the
re-radiated IR radiation and (2) the ``optically-thick'' limit, in
which the disk is optically-thick to the re-radiated IR.

We begin by describing the properties of our models common to both
limits.  We make the clear oversimplification that the disk is a
single phase medium (see, however, the brief discussions in
\S\ref{section:1scat} and Appendix \ref{appendix:stability}); our model
thus describes only the average properties of the disk.  We assume
that the disk is in radial centrifugal balance with a rotation rate
$\Omega=\Omega_{\rm K}$, where $\Omega_{\rm K}=\sqrt{2}\sigma/r$ is
the Keplerian angular frequency in an isothermal potential with
velocity dispersion $\sigma$.  The total dynamical mass at radius $r$
is given by $\mtot(r) =2 \sigma^2 r/G$ and the associated surface
density is \be \stot=\frac{\sigma^2}{\pi G
r}\sim0.6\,\sigma_{200}^2\,\rkpc^{-1}\,\,{\rm g\,\,cm^{-2}},
\label{stot}
\ee where $\sigma_{200} = \sigma/200 \, \kms$ and the radial scale is
in units of kpc. For simplicity, we assume that the underlying
potential is spherical, as would be provided by a stellar bulge or the
galaxy's dark matter halo.\footnote{For a Navarro, Frenk, \& White
(1997) dark matter profile, the dark matter mass can be significant
even on the small scales of interest: $M(r) \approx 10^{10} \, M_\odot
\rkpc^2$; this result is relatively independent of halo mass so long
as $r \ll r_s$, where $r_s$ is the scale radius of the NFW potential.}
We further assume that the gas mass is a constant fraction
$f_g=M_g/\mtot = \Sigma_g/\stot$ of the total dynamical mass, where
$\Sigma_g$ is the gas surface density.  Although the assumption of a
constant $f_g$ is clearly simplistic, we believe that this model
captures much of the physics of interest on large scales.  In
\S\ref{section:disk} we relax this assumption and show that a constant
gas fraction is equivalent to a constant accretion rate.

For a thin disk in a spherical potential, the equation for vertical
hydrostatic equilibrium, $\p p/\p z=-\rho\Omega^2z$, can be
approximated as \be p \approx \rho h^2\Omega^2,
\label{hydrostat}
\ee where $h$ is the pressure scale height.  This implies $h \approx
c_s/\Omega$ where $c_s^2 = p/\rho$ is the sound speed.  Throughout this
paper we include turbulent pressure in our definition of the sound
speed (in addition to radiation pressure and gas pressure). A term
$2\pi G\Sigma_g\rho h$ should be added to the right-hand-side of
equation (\ref{hydrostat}) to account for the self-gravity of the disk.
This leads to a multiplicative correction to equation (\ref{hydrostat})
of $(1+2^{3/2}/Q)$, which we neglect for simplicity.

We assume that star formation in the disk is governed by local
gravitational instability as described by Toomre's Q-parameter (Toomre 1964).  
In particular, we argue that the disk self-regulates such that
\be Q=\frac{\kappa_\Omega c_s}{\pi G
\Sigma_g}=\frac{\Omega^2}{\sqrt{2}\pi G\rho},
\label{q}
\ee is maintained close to unity.  In equation (\ref{q}),
$\kappa^2_\Omega=4\Omega^2+d\Omega^2/d\ln r$ is the epicyclic
frequency.  The hypothesis of marginal Toomre stability has been
discussed extensively in the literature (e.g., Paczynski 1978; Gammie
2001; Sirko \& Goodman 2003; Levin 2005) and is based on the idea that
if $Q \gg 1$ then the disk will cool rapidly and form stars, while if
$Q \ll 1$ then the star formation will be so efficient that the disk
will heat up to $Q \sim 1$.  There is evidence for $Q \sim 1$ in the
Milky Way (e.g., Binney \& Tremaine 1987), local spiral galaxies
(e.g., Martin \& Kennicutt 2001), and starbursts such as ULIRGs (e.g.,
Downes \& Solomon 1998).

From equation (\ref{q}) it follows that the density
distribution of the gas is determined solely by the local Keplerian
frequency: \be \rho=\frac{\Omega^2}{\sqrt{2}\pi G
Q}=\frac{\sqrt{2}\sigma^2}{\pi G Q
r^2}\Longrightarrow n\sim170\sigma_{200}^2\rkpc^{-2}Q^{-1}\,{\rm cm^{-3}}.
\label{rhoq}
\ee 
We retain the $Q$ dependence here and below for completeness, but our
assumption is that $Q\sim1$. From the definition of $\Sigma_g$ and
equations (\ref{stot}) and (\ref{rhoq}), it then follows that \be
(h/r)=f_g Q/2^{3/2}
\label{hzero}
\ee and that 
\be 
(c_s/\sigma)=f_g Q/2.
\label{cszero}
\ee 
For constant $f_g$ and $Q$, $h/r$ and $c_s/\sigma$ are independent
of radius.  With
$\sigma=200$ km s$^{-1}$, a turbulent sound speed of $10$ km s$^{-1}$
corresponds to a gas fraction of $\fg=f_g/0.1$.
The scale height of the disk for this gas fraction is then
\be 
h\sim35\fg\rkpc\,\,{\rm pc}.
\label{scaleh}
\ee 

In our model, the star formation rate per unit area $\dot
\Sigma_\star$ adjusts to maintain $Q \sim 1$.  We also parameterize
star formation as occurring on a fraction $\eta$ of the local dynamical
timescale (e.g., Elmegreen 1997): \be \sds=\Sigma_g \Omega \eta.
\label{sdsform} 
\ee  
Only values of $\eta \lesssim 1$ are physical.  For $\eta$ greater than unity
the disk cannot dynamically adjust to maintain $Q \sim 1$.  
Although, $\dot \Sigma_\star$ is the fundamental derived quantity, 
$\eta$ is a useful alternative parameter that characterizes the
global star formation efficiency in the disk.  Observationally, $\eta$
is typically $\sim0.02$ in normal spiral and starburst galaxies (Kennicutt 1998).


\subsection{The Optically-Thin Limit}
\label{section:1scat}

We consider galactic disks to be ``optically-thin'' when the vertical
optical depth to IR photons is $\lesssim 1$, but the optical depth to
UV photons is greater than unity.  This requires
$\Sigma_g\gtrsim2/\kappa_{\rm UV}\sim10^{-3}$ g cm$^{-2}$, where
$\kappa_{\rm UV}\sim10^3$ cm$^2$ g$^{-1}$ is a characteristic UV
opacity.

In our models, feedback from star formation provides the pressure
support to maintain $Q \sim 1$.  Sources of this pressure support
include radiation pressure on dust grains, supernovae, stellar winds,
and expanding HII regions.  In this section we consider a simple model
that relates the pressure in the ISM to the star formation rate $\sds$
in the optically thin limit (eq. [\ref{1scatter}] below).  To motivate
this model, and to connect with classic treatments of the ISM (e.g.,
McKee \& Ostriker 1977), we estimate the pressure in both the ``cold''
($p_c$) and ``hot'' ($p_h$) phases of the ISM, focusing on the
contributions from radiation pressure and supernovae.  The pressure in
the cold ISM can thus be written as: \be p_c=\rho c_s^2\sim p_{\rm
rp}+p_{\rm sn},
\label{totalpthin}
\ee where $c_s$ should be interpreted as a turbulent velocity and
$p_c$ as a turbulent pressure.

Because we assume that the optical depth to UV photons is greater than
unity, the radiation pressure on cold gas can be related to the star
formation rate by \be p_{\rm rp}\sim\epsilon\sds c
\label{prp}
\ee where $\epsilon$ is the efficiency with which star formation
converts mass into radiation ($\epsilon \sim 10^{-3}$ for a Salpeter
IMF from $1-100 \, M_\odot$).  Equation (\ref{prp}) assumes that there
is no significant cancellation of oppositely directed momentum.  
This becomes an increasingly better assumption as the disk becomes
optically thick to the IR (as in starburst galaxies).  In addition to
radiation pressure, supernova explosions deposit momentum into the
ISM via swept-up shells of cold gas.  From Thornton et al.~(1998) we
estimate that each supernova has an asymptotic momentum of $3 \times
10^{43} E_{51}^{13/14} n_1^{-1/4}$ g cm s$^{-1}$, where
$E_{51}=E/10^{51}\,{\rm ergs}$ is the initial energy of the supernova
and $n_1=n/1\,{\rm cm^{-3}}$ is the density of the ambient medium.
The contribution to the ISM pressure from supernovae may be estimated 
by taking $p_{\rm sn}=\dot{P}_{\rm sn}/(2 \pi r^2)$, 
where $\dot{P}_{\rm sn}$ is the total momentum injection rate.
Alternatively, one can estimate $p_{\rm sn}$ by balancing the
net volumetric energy injection rate by supernovae,
$q_{\rm sn}^+\sim\dot{P}_{\rm sn}c_s/(2\pi r^2 h)$, 
with energy losses due to turbulent decay.
Assuming that turbulence decays on a crossing time
so that $q_{\rm sn}^-\sim \rho c_s^3/h$ (e.g., Stone et al.~1998), 
we see that $q_{\rm sn}^+=q_{\rm sn}^-$
implies that $\dot{P}_{\rm sn}/(2\pi r^2)=\rho c_s^2=p_{\rm sn}$.
Thus, the contribution to the pressure of the cold ISM from supernovae is
\be p_{\rm sn} \sim 1.5 \times
10^8 n_1^{-1/4} E_{51}^{13/14} \, \sds \sim 5 \, n_1^{-1/4}
E_{51}^{13/14} p_{\rm rp}
\label{psn} \ee 
Equations (\ref{prp}) and
(\ref{psn}) show that radiation pressure and supernovae contribute
comparably to the pressure of the cold ISM.

Supernovae also generate a shocked hot ISM, whose pressure $p_{\rm h}$
can be estimated using the model of McKee \& Ostriker (1977).
Individual supernovae initially expand adiabatically 
to a cooling radius $R_c\sim20 E_{51}^{0.3}n_1^{-0.4}$ pc, where a
cool dense shell forms (Chevalier 1974; Cioffi et al.~1985; Thornton
et al.~1998).  The subsequent remnant evolution is approximately
momentum-conserving.  In the absence of interaction with neighboring
supernova remnants, the remnant expands to a maximum radius
$R_p\sim60E_{51}^{0.3}n_1^{-0.2}P_{-12}^{-0.2}$ pc, where the pressure
of the supernova remnant is equal to the ambient pressure ($P$) and
$P_{-12}=P/10^{-12}$ ergs cm$^{-3}$.  In fact, the remnant typically
does not reach $R_p$ before encountering a neighboring supernova
remnant.  That is, the overlap radius for supernova remnants ($R_{\rm
o}$) is $R_c<R_o<R_p$.  In this model, the average pressure of the hot
ISM is not known {\it a priori}, but can be estimated using the
pressure of a supernova remnant at $R_o$ (this assumes that there is
no significant energy loss after the supernova remnants overlap and
mix).  Using results from Chevalier (1974), McKee \& Ostriker (1977)
showed that the pressure at overlap can be expressed in terms of the
pressure at $R_c$ as\footnote{More carefully, $\Psi_c$ should be
raised to the 10/11-th power in equation (\ref{psic}), which we
approximate as unity for simplicity.} \be p_{\rm h}\sim P_c \Psi_c,
\label{psic} \ee where $\Psi_c=SV_c t_c$ is the the number of
supernovae that occur in the volume $V_c\sim(4/3)\pi R_c^3$, $t_c$ is
the time for the remnant to reach $R_c$, and $S$ is the supernova rate
per unit volume.  The time $t_c$ has been estimated by Cox (1972) and
Cioffi et al.~(1985): $t_c\sim4\times10^4 E_{51}^{3/14}n_1^{-4/7}$ yr
for solar metallicity.  Using $P_c=E/V_c$ we find that \be p_{\rm
h}\sim10^{-12}E_{51}^{17/14}n_{-2}^{-4/7}S_{-13}\,\,{\rm
ergs\,\,\,\,cm^{-3}},
\label{phot}
\ee where we have scaled the number density to a value comparable to
that of the hot ISM 
($n_{-2}=n/10^{-2}$ cm$^{-3}$), and the supernova rate to the Galactic
value: $S_{-13}=S/10^{-13}$ pc$^{-3}$ yr$^{-1}$.  Equation
(\ref{phot}) is in reasonable agreement with observations and more detailed
models of the local pressure of the ISM (McKee \& Ostriker 1977;
Boulares \& Cox 1990).

In order to compare $p_{\rm h}$ and $p_{\rm rp}$ from equation
(\ref{prp}) directly, we note that $S\sim10^{-2}\sds/(2h)$ with $\sds$
in $M_\odot$ yr$^{-1}$ per unit area.  The ratio of these two
components of the pressure can then be written as \be {p_{\rm h} \over
p_{\rm rp}}\sim3\,\,h^{-1}_{100}\,n_{-2}^{-4/7}E_{51}^{17/14},
\label{pratio}
\ee where $h_{100}=h/100$ pc.  
Because the total volume occupied by
remnants with $R\le R_c$ decreases as the density of the ISM
increases, the contribution from supernovae to the total pressure
decreases with increasing density.  
Equations (\ref{totalpthin})-(\ref{pratio}) show that, to order of
magnitude, $p_c = p_{\rm rp} + p_{\rm sn} \sim p_h$ for conditions
appropriate to the Galaxy (as is observed).

In the luminous starbursts we focus on in this paper, the density of
the ISM is much larger than in the local ISM.  For example, in the
inner few hundred parsecs of ULIRGs the {\it average} gas density
reaches $10^3-10^4$ cm$^{-3}$, comparable to the density of a local
molecular cloud (e.g., Downes \& Solomon 1998; see also
eq.~[\ref{rhoq}]).  Several lines of evidence suggest that the cold
molecular gas may fill a significant fraction of the volume in ULIRGs,
unlike in the local ISM (Downes, Solomon, \& Radford 1993; Solomon et
al.~1997).  The high fraction and high luminosity of radio supernovae
in Arp 220 is also consistent with an environment much denser than the
ISM of normal spiral galaxies (Smith et al.~1998).  Taking $n \sim
10^3$ cm$^{-3}$ as a characteristic value, we find that $R_c$ is just
$\sim 1$ pc, $p_{\rm h}/p_{\rm rp} \sim 10^{-2}h^{-1}_{10}$, and that
the total asymptotic thermal energy of a supernova remnant is
$\sim4\times10^{48}$ ergs $\ll E$ (Thornton et al. 1998).  This argues
against a dynamically-dominant, volume-filling hot ISM.  Even in the
limit of strong radiative losses, however, supernovae are still
important for generating the random motion of cold gas ($p_{\rm sn}
\sim p_{\rm rp} \gg p_h$; eq.~[\ref{psn}]).

In the simple estimates above, all of the contributions to the
pressure of the ISM scale roughly linearly with the star formation
rate.  Moreover, the ratio $p_{\rm rp}/p_{\rm sn}$ is of order unity,
and while $p_{\rm h}/p_{\rm rp}$ is of order unity in normal spiral
galaxies it may be significantly smaller in the dense nuclei of the
most luminous starbursts.  For this reason we choose to express the
effective pressure of the ISM in the ``optically-thin'' limit as \be p
\sim p_c =p_{\rm rp}[1+(p_{\rm sn}/p_{\rm rp})]=\epsilon\xi\sds c,
\label{1scatter}
\ee where the last equality defines the parameter $\xi$.  In what
follows we retain the dependence of our results on $\xi$, but scale
to $\xi\sim1$ for the reasons given above.

Using equations (\ref{rhoq}), (\ref{cszero}), and (\ref{1scatter}), it
is straightforward to solve for the physical parameters of our disk
model.  The star formation rate per unit area required to support the disk with $Q\sim1$
is given by \be 
\dot\Sigma_\star = {\Sigma_g^2 \Omega^2 \over 4 \rho \epsilon \xi c} =
{\Sigma_g^2 \pi G Q \over 2^{3/2} \epsilon \xi c} = {f_g^2 \sigma^4 Q
\over 2^{3/2} \pi G r^2 \epsilon \xi c}.
\label{sds1scat} 
\ee Scaling equation (\ref{sds1scat}) for typical $f_g$ and $\sigma$
we find that \be \dot\Sigma_\star \sim 35 \fgas^2 \sigma_{200}^4
\rkpc^{-2} \epsilon_{3}^{-1} \xi^{-1} Q \,{\mpykpc} \ee where
$\epsilon_{3} = \epsilon/10^{-3}$ and $\fgas = f_g/0.5$ is appropriate
for gas-rich starbursts.  The star formation rate can also be
expressed in terms of the efficiency $\eta$; \be \eta = {1 \over 2}
\left({c_s \over \epsilon \xi c} \right)= {Q \over 4}\left(\frac{f_g
\sigma }{ \epsilon \xi c}\right) \sim
0.1\fgas\sigma_{200}\epsilon_3^{-1} \xi^{-1}
\label{eta1scat} 
\ee 
The second equality in equation
(\ref{eta1scat}) is meant to show explicitly that one may write $\eta$
in terms of $c_s$ or $f_g\sigma$ (eq.~[\ref{cszero}]).  Using equation
(\ref{sds1scat}), the flux and luminosity of the disk viewed face-on
are given by \be F = \epsilon \dot \Sigma_\star c^2 \sim 5 \times
10^{11} \fgas^2 \sigma_{200}^4 \rkpc^{-2} Q \xi^{-1} \, \lkpc
\label{flux1scat} 
\ee 
and 
\be 
L \equiv \pi r^2 F =
{f_g^2  c Q \over 2^{3/2} G \xi}\sigma^4 \sim 2 \times 10^{12} \fgas^2
\sigma_{200}^4 Q \xi^{-1} \ \, {\rm L_\odot}.
\label{l1scat} 
\ee Up to logarithmic corrections, equation (\ref{l1scat}) implies
that for constant $f_g$ all radii contribute equally to the total
luminosity.

The effective temperature is defined by the relation $\sigma_{\rm
SB}\teff^4=\epsilon\sds c^2$.  In the optically-thin limit, the
observed dust temperature $T_{\rm dust}$ is related to $\teff$ by
$T_{\rm dust}^4\sim\teff^4/(2\tau_{\rm V})$, where $\tau_{\rm V}$ is
the vertical optical depth to IR radiation.  This relation between
$T_{\rm dust}$ and $T_{\rm eff}$ assumes that the sources of UV
radiation are uniformly distributed vertically throughout the disk.
With equation (\ref{sds1scat}), \be \hspace{-0.4cm} T_{\rm
dust}=\left(\frac{f_g\sigma^2}{r}\frac{Qc}{2^{3/2} \kappa \xi
\sigma_{\rm SB}}\right)^{1/4}
\hspace{-0.45cm}
\sim60\fgas^{1/4}\sigma_{200}^{1/2}(\kappa_1 \xi \rkpc)^{-1/4}
Q^{1/4}\,{\rm K},
\label{tdust1}
\ee where $\kappa_1 = \kappa/1 \, {\rm g \, cm^{-2}}$ is a
representative value for the IR opacity of the disk
(Fig.~\ref{plot:kp}) and we have assumed that an individual dust grain
radiates as a blackbody.

There are several interesting properties of the disk model presented
in this section.  First, the efficiency of star formation required to
maintain $Q \sim 1$ is $\eta \sim 0.02$ for a canonical turbulent
velocity of $c_s \sim 10 \, \kms$ (equivalent in our model to having
$f_g \sim 0.1$ for $\sigma \sim 200$ km s $^{-1}$, as in the Galaxy).
This value for $\eta$ is in good agreement with observations compiled
by Kennicutt (1998).
Second, the first two equalities in equation (\ref{sds1scat}) yield
$\dot \Sigma_\star \propto \Sigma_g^2$.  This Schmidt-like star
formation law is somewhat steeper than the $\dot \Sigma_\star \propto
\Sigma_g^{1.4}$ favored by Kennicutt (1998), but comparable to the
scaling $\dot \Sigma_\star \propto \Sigma_g^{1.75}$ obtained by Gao \&
Solomon (2004) using a sample that includes more luminous starburst
galaxies.  Given the simplicity of the model presented here, this
agreement is satisfactory.  Third, for $f_g=0.1$ and $\sigma=200$ km
s$^{-1}$ the dust temperature (eq.~[\ref{tdust1}]), turbulent
velocity, pressure, flux, luminosity, and scale height
(eq.~[\ref{scaleh}]; compare with Fig.~9.25 from Binney \& Merrifield
1998) are all in fair agreement with observations of the Milky Way.
Thus despite its simplicity, the model presented in this section
provides a useful characterization of galactic-scale star formation
supported by the turbulent pressure produced by supernovae and the
radiation from massive stars.


\subsection{The Optically-Thick Limit}
\label{section:diffusive}

The nuclei of gas-rich starbursts are optically thick to their own
infrared radiation.  Radiative diffusion then ensures that radiation
pressure provides the dominant vertical support against gravity.  In
this section we describe disk models appropriate to this limit.  The
vertical optical depth of the disk is given by $\tauv = \Sigma_g
\kappa/2$, where $\kappa$ is the Rosseland mean opacity to dust.
Evaluating this expression yields \be
\tauv=\frac{\kappa\sigma^2f_g}{2\pi G
r}\sim0.15\sigma_{200}^2\fgas\rkpc^{-1}\kappa_1.
\label{tauzero}
\ee 
The radius at which $\tauv = 1$ is then
\be 
\rtauv=\frac{\kappa\sigma^2f_g}{2\pi
G}\simeq150\kappa_1\sigma_{200}^2\fgas\,{\rm pc}
\label{rtauv} 
\ee 
and thus for the largest most gas-rich starbursts ($\sigma\sim300$ km s$^{-1}$ 
and $f_g\sim1$) the inner $\sim700$ pc are optically thick.

In the optically thick limit, the effective temperature is given by
\be \sigma_{\rm SB} T_{\rm eff}^4 =\frac{1}{2}\epsilon \sds c^2,
\label{teff1} 
\ee 
where the factor of $1/2$ arises because
both the top and bottom surfaces of the disk radiate.  
The midplane temperature is related to the effective temperature by
$T^4\approx(3/4)\tauv\teff^4$, where the opacity $\kappa(T,\rho)$ in
$\tauv$ should be evaluated using the central temperature and mass
density of the disk.  The temperature dependence of the 
opacity is important for this problem and is discussed in the next
section.  For now we simply normalize $\kappa$ to 1 cm$^2$ g$^{-1}$.

For $\tauv \gtrsim 1$, the radiation pressure is given by \be p_{\rm
rad} = {4\ssb \over 3c} { T^4} = {\ssb \over c}\tauv \teff^4 = {1
\over 2} \tauv \epsilon {\dot \Sigma_\star} c.
\label{prad} 
\ee Comparing equations (\ref{1scatter}) and (\ref{prad}) shows that
radiation pressure exceeds the turbulent pressure due to supernovae by
a factor of $\sim \tauv$ (assuming $\xi \sim 1$ for the reasons given
in \S\ref{section:1scat}).  Radiation pressure support will thus
dominate the vertical support of compact optically-thick starbursts.
The exact surface density (or $\tauv$) at which the transition to
radiation pressure support occurs is somewhat uncertain and requires a
better understanding of the pressure support provided by supernovae,
stellar winds, etc..  We will explore this in more detail in future
work.

With equation (\ref{prad}) it is again straightforward to solve for
the disk's physical parameters, $T$, $T_{\rm eff}$, $\dot
\Sigma_\star$, etc.  in terms of our model variables: $\sigma$, $f_g$,
$\epsilon$, and the radius in the disk $r$.  The midplane temperature
of the disk is \be
T=\left(\frac{f_g\sigma^2}{r}\right)^{1/2}\hspace{-.15cm}\left(\frac{3cQ}{2^{7/2}\pi
G \ssb}\right)^{1/4}
\hspace{-.45cm}
\sim 41\,\sigma_{200}\fgas^{1/2}\rkpc^{-1/2}Q^{1/4}\,{\rm K}.
\label{temp}
\ee
and the effective temperature is 
\be
\teff=\left(\frac{f_g\sigma^2}{r}\frac{cQ}{\sqrt{2}\kappa\ssb}\right)^{1/4}
\hspace{-.45cm}
\sim 70\,
\sigma_{200}^{1/2}\fgas^{1/4}\rkpc^{-1/4}\kappa_1^{-1/4}Q^{1/4}\,{\rm
K}. \label{teff} 
\ee 
For our fiducial numbers, $\teff$ is somewhat
larger than $T$, implying that the vertical optical depth is less than
unity.  Our assumption that the disk is optically thick in the far
IR is therefore only marginally applicable on kpc scales.  It
is an increasingly better assumption on smaller scales (see
eqs.~[\ref{tauzero}] and [\ref{rtauv}]).

The total star formation rate per unit area required to support the disk
with radiation pressure is
\be
\sds=\frac{\sqrt{2}f_gQ}{\epsilon\kappa c}\frac{\sigma^2}{r} 
\sim 
400 \fgas\sigma_{200}^2Q\rkpc^{-1}\kappa_1^{-1}\epsilon_3^{-1}\,{\rm
M_\odot\,\,yr^{-1}\,\,kpc^{-2}}.
\label{sds}
\ee
Integrating, we derive the total star formation rate:
\be
\mds=\frac{2^{3/2}\pi f_g Q \sigma^2
r}{\epsilon\kappa c}
\sim
3000\fgas\sigma_{200}^2\rkpc
Q\epsilon_3^{-1}\kappa_1^{-1}\,{\rm M_\odot \,\,yr^{-1}}. 
\label{sfr}
\ee
Note that if the disk is optically-thick ($2\times\tauv\sim1$),
the observed flux is only from half of the disk and therefore the
observationally inferred star formation rate would be one-half of the
true total star formation rate (given in eqs. [\ref{sds}] \& [\ref{sfr}]).

The star formation efficiency $\eta$ is
\be 
\eta=\frac{\pi G Q}{\kappa\epsilon
c}\frac{r}{\sigma} \sim 1 \,\, \rkpc\sigma_{200}^{-1}
\epsilon_3^{-1}\kappa_1^{-1}Q.
\label{eta}
\ee 
For $r\lesssim R_{\tau_{\rm V}=1}$ we see that $\eta$ is $<1$.
It follows that the disk can adjust
to maintain $Q \sim 1$ on sub-kpc scales.  In addition, we find that
the ratio of the cooling timescale to the orbital timescale is much
less than unity, so the disk should self-regulate to maintain $Q \sim
1$ (see Appendix \ref{appendix:stability}).  

Finally, the surface brightness of the
disk viewed face-on is 
\be
F=\frac{f_gQc}{\sqrt{2}\kappa}\frac{\sigma^2}{r}
\sim3\times10^{12}
\fgas\sigma_{200}^2\rkpc^{-1}Q\kappa_1^{-1}\,\,{\rm
L_\odot\,\,kpc^{-2}} \label{ftot} 
\ee
and the total luminosity for a single side of the disk is 
\be
L=
\frac{\sqrt{2}\pi f_g Q c
r\sigma^2 }{\kappa } \sim2\times10^{13}
\fgas\sigma_{200}^2\rkpc Q \kappa_1^{-1}\,\,{\rm
L_\odot}. 
\label{ltot} 
\ee

Equations (\ref{teff})-(\ref{ltot}) show that the observable
properties of the disk depend sensitively on the magnitude of the
opacity.  In particular, the star formation rate per unit area is
proportional to $\Sigma_g^2/\tau_{\rm V}\sim\Sigma_g/\kappa$
(eq.~[\ref{sds}]).  Therefore, in regions of the disk where the
opacity is low, there must be more star formation to maintain $Q\sim1$.  
Conversely, where the
optical depth is high, less star formation is required.  The
functional dependence of $\sds$ in the optically-thick limit should
compared with the equation (\ref{sds1scat}), which shows that in the
optically-thin limit $\sds\propto\Sigma_g^2$.  Therefore, the
``Schmidt-law'' for star formation changes in the dense
optically-thick inner regions of starburst galaxies.

Equation (\ref{ftot}) can be rewritten as 
\be 
\frac{L}{M}={F \over \Sigma_{\rm tot}} =
{\pi Q G c f_g \over \sqrt{2} \kappa} = {2 \pi G c \over \kappa}{h
\over r} \sim 10^3 \, {\fgas \, Q \over \kappa_1} \, 
{\rm L_\odot \over M_\odot}. 
\label{edd} 
\ee 
The third equality above is the classical Eddington limit,
modified by the factor $(h/r)\propto f_g$, and with the Rosseland mean
opacity taking the place of the usual electron scattering opacity.\footnote{The factor $2 \pi$
rather than the more standard $4 \pi$ in equation (\ref{edd}) is
because the luminosity in equation (\ref{ltot}) is only from 1/2 of
the disk.}  This way of expressing the flux highlights the fact that
each disk annulus radiates at its local Eddington limit.  The value
derived here for the luminosity per unit mass from the disk is similar
to that estimated by Scoville (2003), who argues that this limit is
observed in both young star clusters such as M51 and ULIRGs such as
Arp 220 (see also Scoville et al.~2001).  
Note that for a given gas fraction, the mass-to-light ratio is
proportional to the dust opacity and is therefore metallicity dependent.


\subsection{Opacity Dependence}
\label{section:opacitytemp}

\begin{figure}
\begin{center}
\plotone{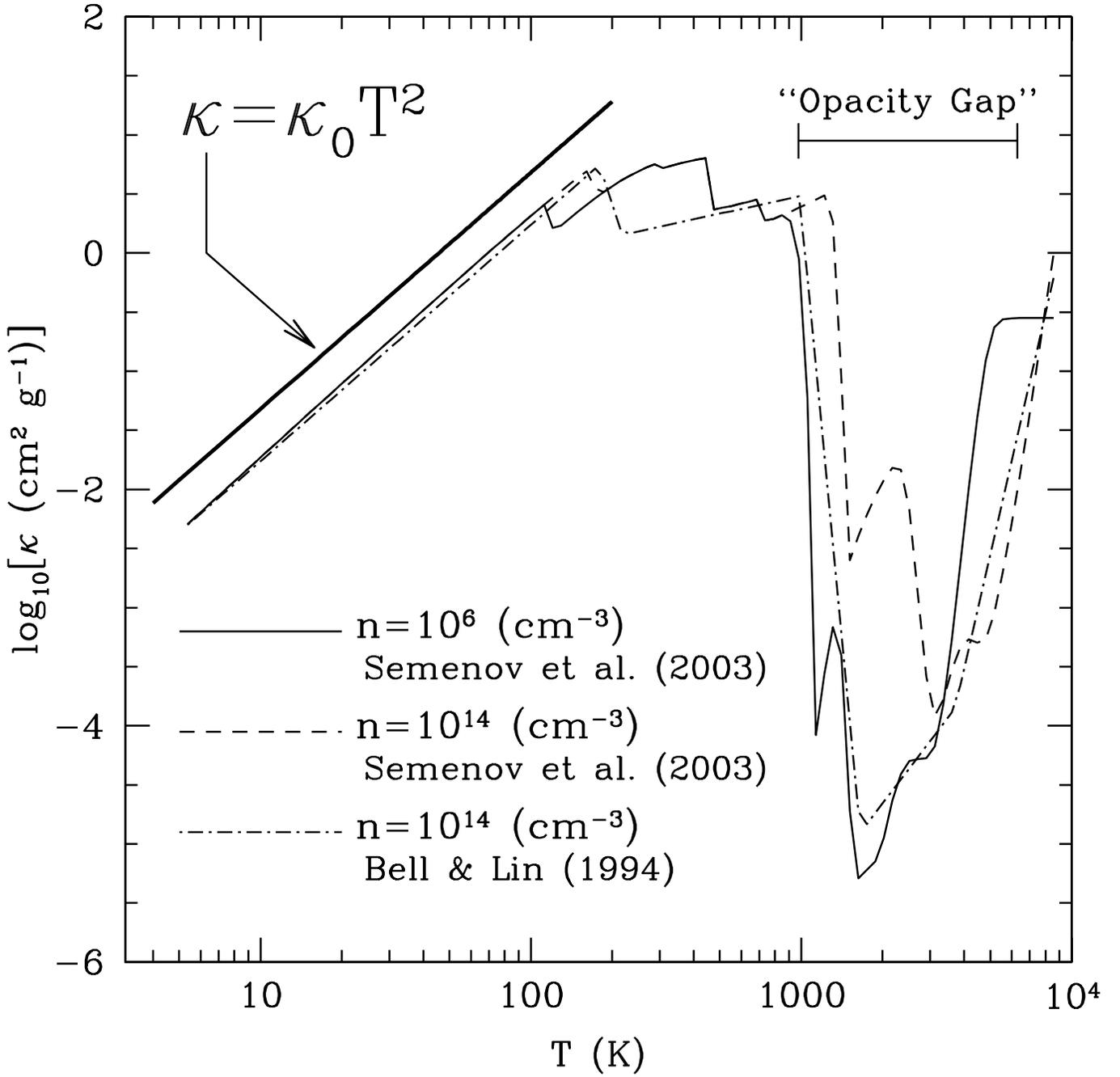}
\end{center}
\figcaption{The Rosseland mean opacity as a function of temperature at
several densities using two opacity models from the literature.  Note
the pronounced decrease in the opacity at the sublimation temperature
of dust at $T\sim1000$ K (the ``opacity gap'').  The rise in the
opacity at higher temperature is due first to H-scattering, then to
bound-free and free-free interactions, and finally to scattering off
of free electrons.  The thick solid line shows that at low
temperatures, the opacity scales with the square of the temperature. 
\label{plot:kp}}
\end{figure}

Figure \ref{plot:kp} shows the Rosseland mean opacity as a function of
temperature for several densities using the publicly available
opacities of Semenov et al.~(2003) and Bell \& Lin (1994).  There are
two important features.  First, for temperatures $T \lesssim 100-200$
K, the opacity is essentially independent of density and can be
approximated by $\kappa=\kappa_0 T^{2}$, with
$\kappa_0\simeq2.4\times10^{-4}$ cm$^2$ g$^{-1}$ K$^{-2}$. The scaling
of $\kappa$ with $T^2$ follows from the fact that the dust absorption
cross section scales as $\lambda^{-\delta}$ with $\delta\rightarrow2$
in the Rayleigh limit (Pollack et al.~1985).  The normalization
$\kappa_0$ is somewhat uncertain and depends on grain physics and the
dust-to-gas ratio; the latter may vary systematically as a function of
radius and metallicity in starburst disks.  In fact, in our own galaxy
there is evidence that the dust-to-gas ratio increases within the
central few kpc (Sodroski et al.~1997).  In what follows we
set $\kappa_{-3.6}=\kappa_0/2.4\times10^{-4}$ cm$^2$ g$^{-1}$ K$^{-2}$
and retain the scaling with $\kappa_0$.  The second important feature
of Figure \ref{plot:kp} is the dramatic decrease in the opacity for
$10^3 \, {\rm K} \lesssim T \lesssim 10^4$ K, the ``opacity gap.''
Here, the temperature is larger than the sublimation temperature of
dust, but smaller than the temperature at which hydrogen is
significantly ionized.

Equation (\ref{temp}) shows that, even for the largest galaxies with
the highest gas fractions, the temperature at $\sim 0.1-1$ kpc is
$\lesssim100$ K, and so the opacity on large scales can be approximated by 
$\kappa =\kappa_0 T^2$.  We may then eliminate the opacity
dependence from the disk properties derived above.  Remarkably, because
$T\propto\Sigma_g^{1/2}$ and $\sds\propto\Sigma_g/\kappa$, with
$\kappa\propto T^2$ we find that the star formation rate per unit
area, the effective temperature, and flux are all independent of
virtually all model parameters: 
\be
\hspace*{-.1cm}\sds=\left(\frac{2^{9/2}\pi G \ssb Q}{3\epsilon^2\kappa_0^2c^3}\right)^{1/2}
\hspace{-.4cm}\sim10^3\,\epsilon_3^{-1}\kappa_{-3.6}^{-1}Q^{1/2}\,\,{\rm M_\odot\,yr^{-1}\,kpc^{-2}}
\label{sdsopacity}
\ee
\be
\hspace*{-.1cm}{F}=\left(\frac{2^{5/2}\pi G \ssb Qc}{3\kappa_0^2}\right)^{1/2}
\hspace{-.4cm}\sim 10^{13}\,\kappa_{-3.6}^{-1}Q^{1/2}\,\,{\rm L_\odot\,kpc^{-2}},
\label{fluxopacity}
\ee
and
\be \teff=\left(\frac{2^{5/2}\pi G Q
c}{3\kappa_0^2\ssb}\right)^{1/8}\sim88\,\kappa_{-3.6}^{-1/4}Q^{1/8}
\,\,{\rm K}.
\label{teffopacity}
\ee 
In particular, note that
neither $\dot \Sigma_\star$, ${F}$, nor $T_{\rm eff}$ depend on $r$,
$\sigma$, or $f_g$.  Our model thus predicts that starburst disks have roughly
constant flux and effective temperature over a range of radii.  

The constancy of these disk observables follows from three
ingredients: (1) the disk is supported by radiation pressure and
$\tauv \gtrsim 1$, (2) the disk self-regulates with $Q\sim1$, and (3)
$\kappa\propto T^2$ at low $T$.  Above $T\sim100-200$ K, $\kappa$
ceases to increase monotonically with $T$ (see Fig. \ref{plot:kp}) and
equations (\ref{sdsopacity})-(\ref{teffopacity}) no longer hold.
Because $T\propto r^{-1/2}$ (eq.~[\ref{temp}]), the temperature
exceeds $\sim200$ K at a radius $R_{200}\sim40\sigma_{200}^2\fgas
T_{200}^{-2}Q^{1/2}$ pc. This radius should be compared with the
radius at which $\tau_{\rm V}=1$. Using the $\kappa\propto T^2$
scaling, we find that 
\be R_{\tau_{\rm
V}=1}=\sigma^2f_g\left(\frac{3c\kappa_0^2Q}{2^{11/2}\ssb(\pi
G)^3}\right)^{1/4}\hspace{-0.4cm}\sim250\sigma_{200}^2\fgas\kappa_{-3.6}^{1/2}Q^{1/4}{\rm
pc} \ee (compare with eq.~[\ref{rtauv}]).  Therefore, we expect
$\sds$, $\teff$, and $F$ to be roughly constant in the radial range
$R_{200}\lesssim r \lesssim R_{\tau_{\rm V}=1}$, $\sim$ hundreds of
parsecs for fiducial parameters.

For completeness we note that the scaling for the dust temperature in
the optically-thin limit can also be re-written using $\kappa \propto
T^2$ (eq.~[\ref{tdust1}]):
\be 
T_{\rm dust}\sim60\fgas^{1/6}\sigma_{200}^{1/3}\rkpc^{-1/6}\kappa_{-3.6}^{-1/6}Q^{1/6}\xi^{-1/6}\,{\rm
K}.
\label{tdust2}
\ee The weak scaling with model parameters in equation (\ref{tdust2})
implies that the dust temperature should not vary significantly from
system to system, as appears to be observed (e.g., Yun \& Carilli
2002).

\subsection{Combining the Optically-Thin \& Optically-Thick Limits}
\label{section:ssanddiff}

The optically-thin and optically-thick
limits can be combined by expressing the pressure as \be p=\epsilon\sds
c\left(\frac{1}{2}\tau_{\rm V}+\xi\right)
\label{tempint}
\ee using the full temperature-dependent opacity curve
(Fig.~\ref{plot:kp}).  Recall that in the limit $\tauv \gg 1$
equation (\ref{tempint}) describes true radiation pressure support
while in the limit $\tauv \ll 1$ it describes turbulent support with
contributions from supernovae and radiation pressure.
Because the opacity depends on temperature, we must connect the
central temperature with the effective temperature in order to solve
equation (\ref{tempint}).  By interpolating between the optically-thin
and optically-thick regimes we obtain (see also Sirko \& Goodman 2003)
\be T^4=\frac{3}{4}\teff^4\left(\tau_{\rm V}+\frac{2}{3\tau_{\rm
V}}+\frac{4}{3}\right).
\label{tempfunc}
\ee In the optically-thick limit, equations (\ref{tempint}) and
(\ref{tempfunc}) combine to yield $(4\sigma_{\rm SB}/3c)T^4\sim p$,
whereas in the optically-thin limit $(4\sigma_{\rm SB}/3c)2\tau_{\rm
V}T^4\sim p$.  In solving equations (\ref{tempint}) and
(\ref{tempfunc}), we find multiple solutions because of the
complicated temperature dependence of the opacity.  In Appendix
\ref{appendix:stability}, we address the thermal and viscous stability
of these solutions.  We argue that there is a single stable physical
low-temperature solution and focus on this solution throughout the
paper.

\begin{figure}
\begin{center}
\plotone{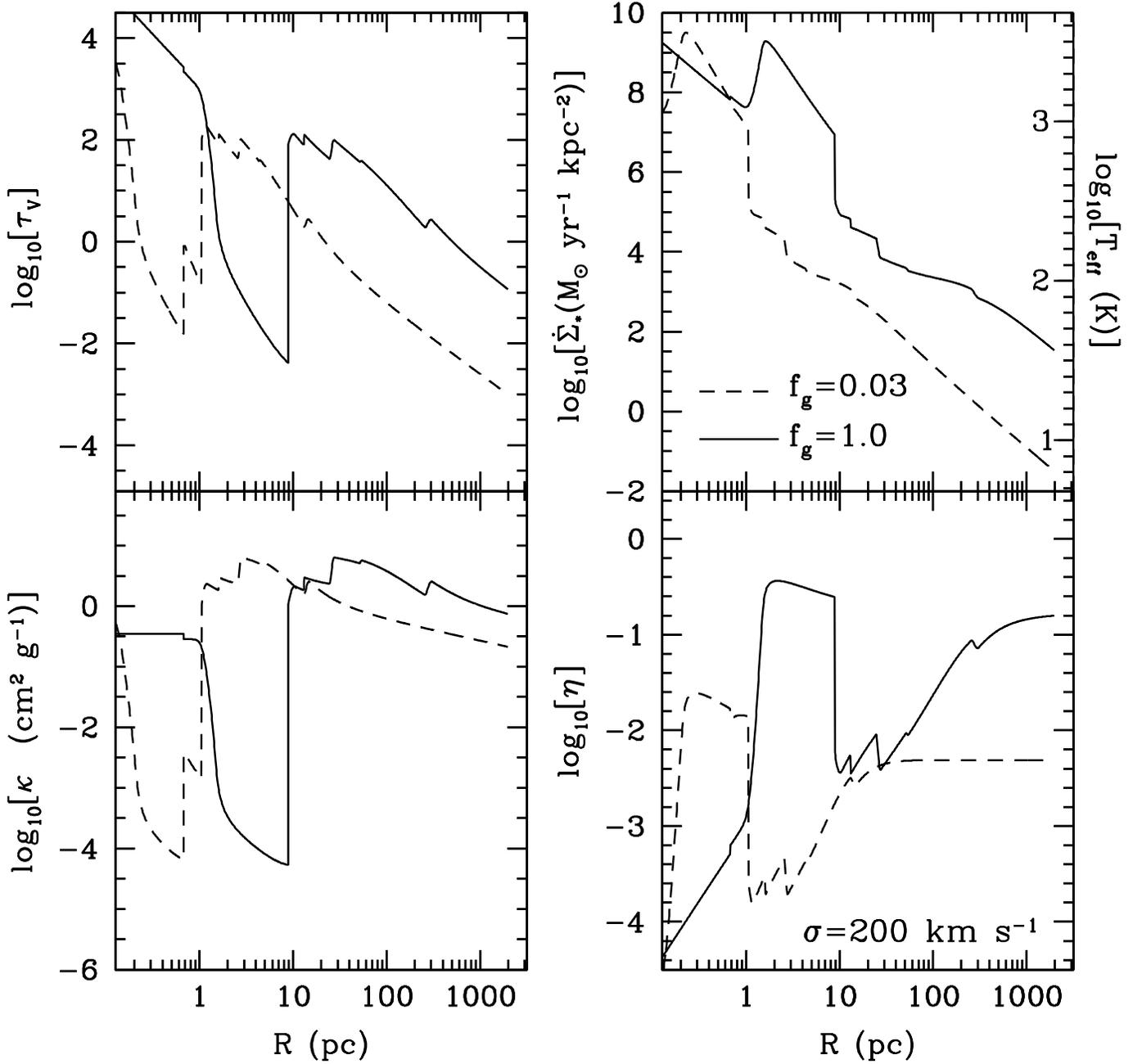}
\end{center}
\figcaption{Vertical optical depth ($\log_{10}[\tauv]$; upper left hand
panel), opacity ($\log_{10}[\kappa\,\,({\rm cm^2\,g^{-1}})]$; lower
left hand panel), star formation efficiency ($\log_{10}[\eta]$;
lower right hand panel), and both the star formation rate per unit
area and the effective temperature ($\log_{10}[\sds\,\,({\rm
M_\odot\,yr^{-1}\,kpc^{-2}})]$, $\log_{10}[\teff\,\,({\rm K})]$; upper
right hand panel) as a function of radius for models with $\sigma =
200 \kms$ and either $f_g=0.03$ (dashed line) or $f_g=1$ (solid line).
$\eta$ approaches a constant for $r \gtrsim 100$ pc in
the model with $f_g=0.03$ when the disk is optically thin (see also
eq. [\ref{eta1scat}]).  The model with $f_g=1$ demonstrates that the
star formation rate per unit area is roughly constant at intermediate
radii $\sim 100$ pc (see \S\ref{section:opacitytemp}).  At small radii
$\sim 0.01-10$ pc, $\sds$ is extremely large in the region where the
dust sublimates and the opacity drops precipitously.  Such large star
formation rates are unphysical since the mass accreting through the
starburst disk would quickly be exhausted.  In \S\ref{section:disk} we
construct more realistic models, taking into account the depletion of
the gas locally as a result of star formation.  Including this effect
significantly reduces $\sds$ on few-parsec scales (compare with
Fig.~\ref{plot:mds}).
\label{plot:disks}}
\end{figure}

Figure \ref{plot:disks} shows the numerically calculated structure of
our disk models for $\sigma = 200 \, \kms$ and with $f_g = 0.03$ and
$1$.  There are three regimes to notice in Figure \ref{plot:disks}.
First, at large radii the disk is optically thin.  In this region
$\dot \Sigma_\star \propto r^{-2}$ and $\eta \sim {\rm const}$ (see \S
\ref{section:1scat}).  At smaller radii the disk becomes optically
thick.  There is then a range of radii ($50\lesssim r \lesssim 300$ pc
for $f_g = 1$) where $\dot \Sigma_\star$ and $T_{\rm eff}$ are roughly
constant, in good agreement with the estimates in
\S\ref{section:opacitytemp}.  Note that where the disk is optically
thin $T^4_{\rm dust}\sim\teff^4/(2\tau_{\rm V})$.  At very small radii
$\sim 1-10$ pc the opacity decreases dramatically when dust grains
sublimate (the ``opacity gap''; Fig.~\ref{plot:kp}).  In this region
the disk becomes optically thin, and the star formation rate required
to maintain $Q \sim 1$ increases significantly (see also Sirko \&
Goodman 2003).  We present a detailed discussion of this part of the
disk in \S\ref{section:disk} and Appendix \ref{appendix:accrete}, but
note here that it is unphysical to assume that the gas fraction is
constant throughout the region where the star formation rate increases
so markedly.

\subsection{Application to ULIRGs}
\label{section:comparisonzero}

To focus the discussion, we emphasize the application of our optically
thick disk models to Arp 220, a prototypical ULIRG.  Arp 220 consists
of two merging nuclei separated by about 350 pc (Graham et al.~1990).
The total FIR luminosity of the system is $\sim 10^{12}$ L$_\odot$.
The $2-10$ keV X-ray luminosity is only $\sim 3 \times 10^9$
L$_\odot$, however, and the column density of X-ray absorbing material
must exceed $10^{25}$ cm$^{-2}$ if an obscured AGN is to contribute
significantly to the bolometric luminosity (Iwasawa et al.~2001).
Thus, there is little evidence for an energetically important AGN.
The detection of numerous radio supernovae also supports a starburst
origin for most of the radiation from Arp 220 (Smith et al. 1998).

The stellar velocity dispersion of Arp 220 is $\sim165$ km s$^{-1}$
(Genzel et al.~2001) and it has  
an extended CO disk with a scale of $\sim 500$ pc (Downes
\& Solomon 1998).  Perhaps $\sim 1/2$ of the luminosity, however,
appears to come from two counter-rotating nuclear disks, each of which
is $\sim 100$ pc in extent (e.g., Sakamoto et al. 1999; Soifer et
al. 1999; Downes \& Solomon 1998).  Radio observations also show that
nearly all of the radio flux --- presumably associated with supernovae
and star formation --- originates within $\sim 50-100$ pc of the
double nuclei (Condon et al.~1991).  The total mass in each of the
compact nuclear disks is $\sim 2 \times 10^9 M_\odot$, with a large
gas fraction of $f_g \sim 0.5$ (Downes \& Solomon 1998).  The nuclear
region of Arp 220 is optically thick to at least $25 \mu$m and the
inferred blackbody temperature at this wavelength is $\gtrsim 85$ K
(Soifer et al. 1999).  The estimated gas mass and radius of the disk
imply a surface density of $\Sigma_g \approx 10$ g cm$^{-2}$,
suggesting that the nuclear region is probably optically thick even in
the FIR (as implied by eq.~[\ref{rtauv}]).

\begin{figure}
\begin{center}
\plotone{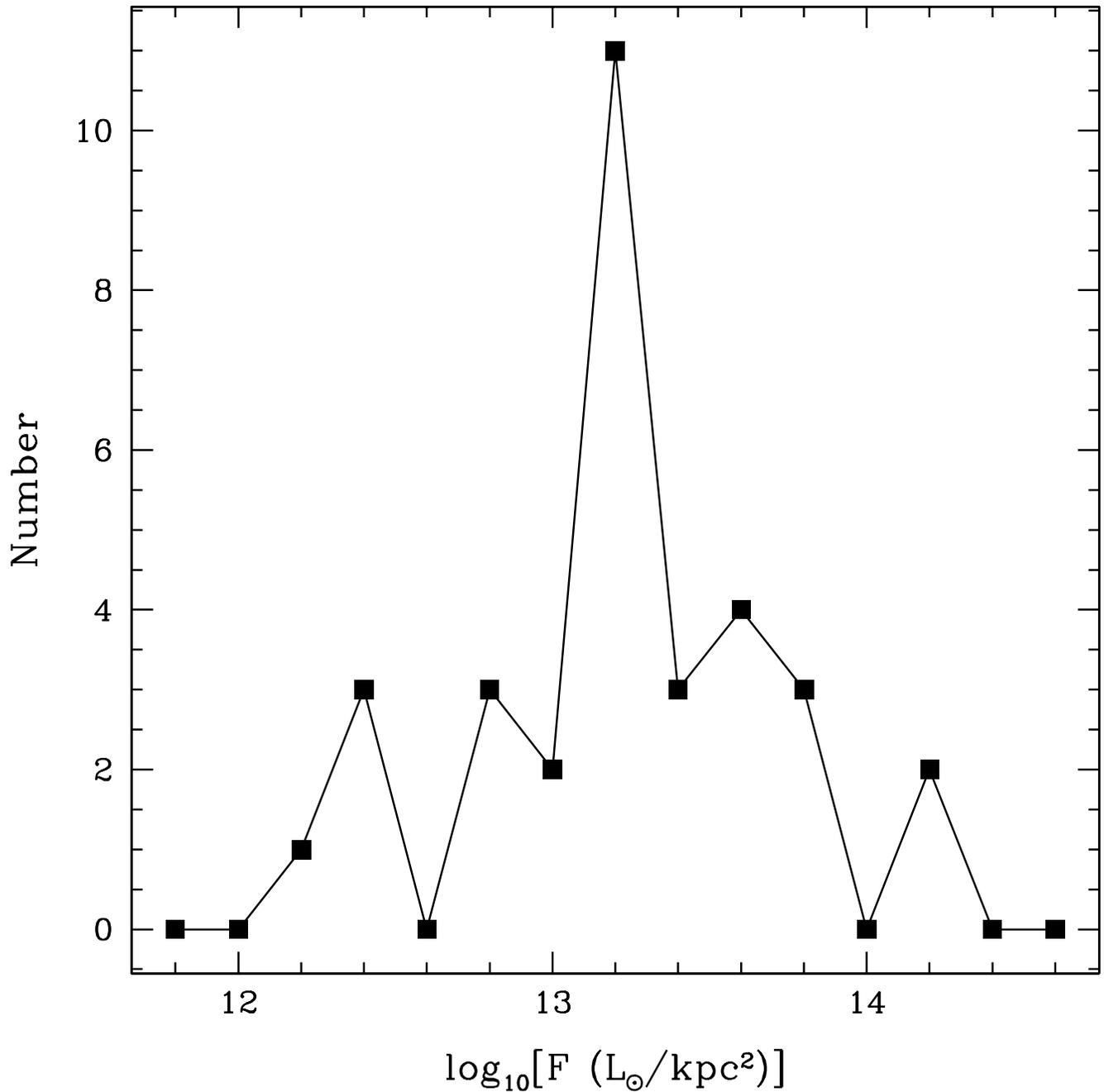}
\end{center}
\figcaption{Histogram of the number of ULIRGs at a given flux using the
sample in Tables 1 \& 2 of Condon et al.~(1991); the bins have a width
of $0.2$ in $\log_{10}{\rm F}$.  We estimate the intrinsic flux in the
nuclear starburst using $F=L_{\rm FIR}/A_{\rm radio}$, where $A_{\rm
radio}$ is the area of Condon et al's elliptical Gaussian fit to the
radio emission; we exclude the 4 unresolved sources.  The peak in the
histogram at $\sim 10^{13} \lkpc$ is in good agreement with the
predictions of our optically thick disk models
(\S\ref{section:opacitytemp}).  See the text in
\S\ref{section:comparisonzero} for uncertainties associated with using
the radio size as a proxy for the starburst size.
\label{plot:fluxdist}}
\end{figure}

In the optically-thick limit, with $\kappa \propto T^2$, our model
predicts a radiative flux of $\sim 10^{13} \, {\rm L_\odot\,kpc^{-2}}$
(eq. [\ref{fluxopacity}]), in good agreement with the flux inferred
for the compact nuclei in Arp 220 (see also Fig. \ref{plot:fluxdist}
discussed below).  Our model also predicts an effective temperature of
$\sim 88$ K for this very compact emission (eq. [\ref{teffopacity}]),
in agreement with that estimated by Soifer et al.~(1999).  Lastly, we
note that for a gas fraction of $f_g \sim 0.5$ and a velocity
dispersion of $\sigma \sim 200 \, \kms$, our model predicts a sound
speed (i.e., turbulent velocity or line-width) of $c_s \sim 50 \,
\kms$, similar to that observed in CO by Downes \& Solomon (1998).

In their study of CO emission from ULIRGs, Downes \& Solomon (1998)
identified compact nuclear starbursts in Mrk 273 and Arp 193 with
properties similar to those discussed above for Arp 220.  More
directly, Condon et al.~(1991) imaged the radio emission in the 40
brightest galaxies in the IRAS Bright Galaxy Sample.  They resolved
the emission in nearly all of the sources (36/40) and found sizes
$\sim 100$ pc.  They further showed that the radio emission from
ULIRGs --- correcting for free-free absorption in some cases --- places
them on the FIR-radio correlation of starburst galaxies.  This is
consistent with starbursts contributing significantly to the
bolometric power of these sources.  In the absence of direct FIR
imaging of ULIRGs, the radio sizes from Condon et al.~(1991) are
currently the best estimate of the size of the nuclear starbursts in
these systems.  Figure \ref{plot:fluxdist} shows a histogram of the
number of sources at a given flux inferred using the FIR luminosity
and the size of the resolved radio source from Condon et al.~(1991).
The histogram shows a peak centered on a flux $\sim 10^{13} \lkpc$.  
This characteristic
`observed' flux is consistent with the predictions of our optically
thick disk models (eq. [\ref{fluxopacity}]).\footnote{Note that our
conclusions are not significantly changed if only (say) $\sim 1/2$ of
the bolometric power is produced by the compact nuclei resolved in the
radio (as some models of Arp 220 suggest; e.g., Soifer et al. 1999).}

\begin{figure}
\begin{center}
\plotone{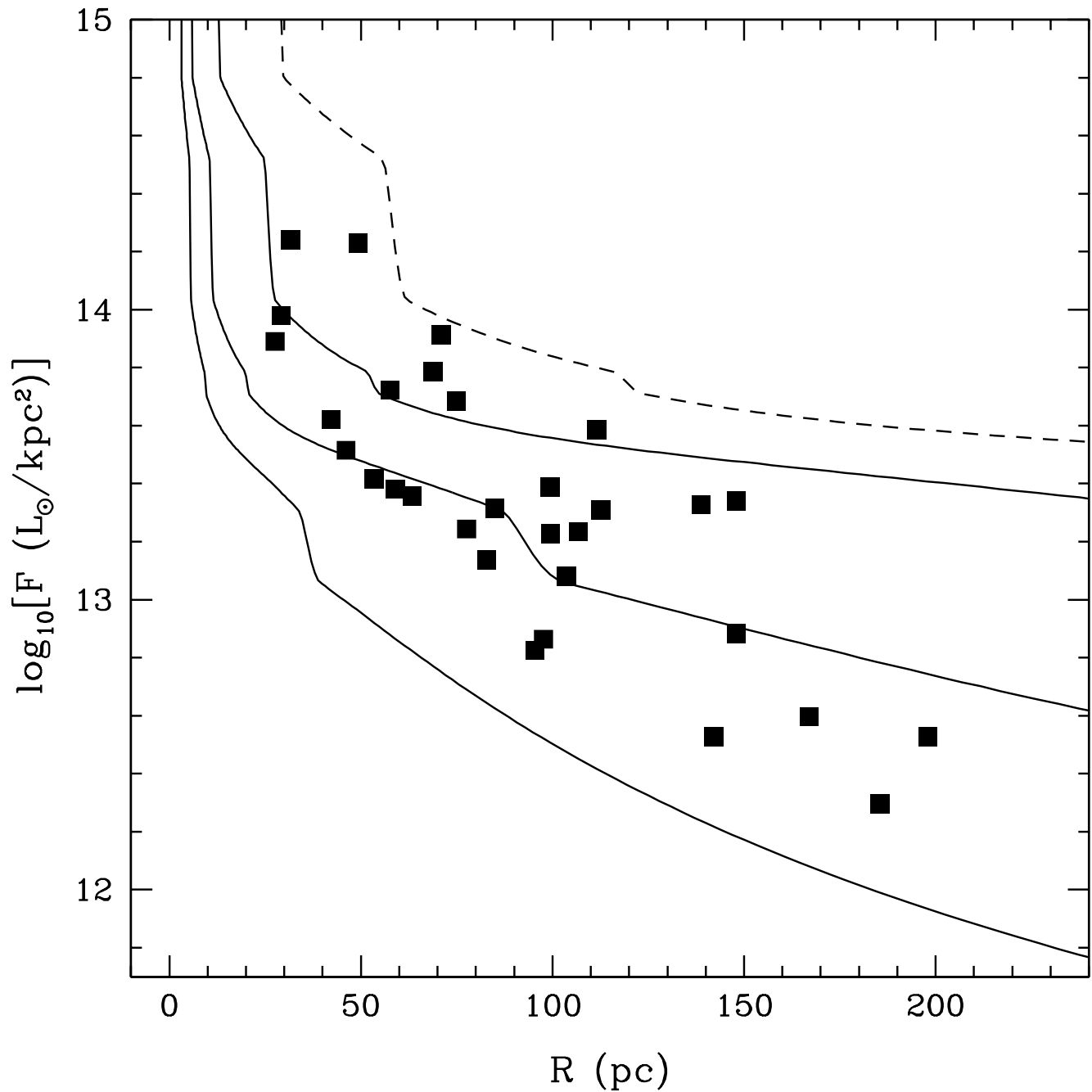}
\end{center}
\figcaption{Flux as a function of radial scale for the local ULIRGs in
Condon et al.~(1991) (filled squares).  The flux is inferred using
$F=L_{\rm FIR}/A_{\rm radio}$, where $A_{\rm radio}$ is the area of
Condon et al's elliptical Gaussian fit to the radio emission; we
define the radial scale $R$ by $A_{\rm radio} \equiv \pi R^2$.  The
solid lines show results from our constant $f_g$ models (compare with
Figs.~\ref{plot:disks}).  From lowest to highest flux the models have
$\sigma=200$ km s$^{-1}$ and $f_g=0.1$, 0.3, and 1.0 (solid lines).
For comparison, the dashed line shows a model computed with
$\sigma=300$ km s$^{-1}$ and $f_g=1.0$.
\label{plot:fluxr}}
\end{figure}

An alternative way to present this data is shown in Figure
\ref{plot:fluxr}, where we plot the inferred flux as a function of the
size of the resolved radio source.  Superimposed on the data we plot
the predictions of our disk models for $\sigma = 200 \kms$ and several
$f_g$ (solid lines), and for $\sigma = 300 \kms$ and $f_g = 1$ (dashed
line).  The latter is a plausible upper limit to the emission expected
in our starburst models.  This Figure demonstrates the excellent
agreement between our models and the observations for $f_g \sim
0.3-1$.  The data are also consistent with the increase in flux we
predict for more compact starbursts.  Taken together, the results of
Figures \ref{plot:fluxdist} and \ref{plot:fluxr} provide strong
support for our interpretation of ULIRGs as Eddington-limited
starbursts.  

It is also worth comparing these results to an empirical surface
brightness limit for starbursts found by Meurer et al. (1997), who
argued that starburst fluxes satisfy $F \lesssim 2 \times 10^{11}
\lkpc$, which corresponds to $\dot \Sigma_\star \lesssim 13 \,
\epsilon_3^{-1} \mpykpc$.  Our calculations, and the observations
summarized above, suggest that Meurer et al.'s limit does not
represent a fundamental limit to the surface brightness of starburst
galaxies.

The characteristic flux $\sim 10^{13} \lkpc$ found in Figure
\ref{plot:fluxdist} is equivalent to a blackbody temperature of $\sim
90$ K (eq. [\ref{teffopacity}]).  This is noticeably larger than the
typical color temperature of $\sim 60$ K inferred from the FIR spectra
of ULIRGs.  Another way to state this result is that using the
observed FIR spectra and luminosity, the blackbody size of the FIR
emitting region is typically larger (by a factor of few) than the
radio sizes observed by Condon et al.~(1991).  This is likely because
the compact nuclei of many ULIRGs are optically thick even at $\sim 30
\mu m$, so that radiative diffusion degrades the $\sim 90$ K emission
and ensures that the FIR size can be larger than the true size of the
nuclear starburst.  This interpretation requires sufficient obscuring
gas at large radii, but also that star formation in this gas does not
dominate the bolometric power of the source (or else the radio source
would be more extended than is observed).  In our models we find that
if the gas fraction in the nuclear region increases at small radii, as
Downes \& Solomon (1998) infer for several systems, then most of the
luminosity is produced near the radius where $\tau_{\rm V} \sim 1$,
rather than in the extended optically thin portion of the disk at
larger radii.

To conclude this section, we compare our model predictions with the
observations of high-redshift ULIRGs (``submm sources'') presented in
Tacconi et al.~(2005).  Tacconi et al. describe CO observations of 4
high-$z$ systems with the PdBI.  They find an average velocity
dispersion of $\sigma\simeq290\pm35$ km s$^{-1}$, gas fraction of $f_g
\simeq0.4\pm0.2$, and half-power radius for CO of $R_{\rm CO} \simeq
1.6\pm0.3$ kpc.  The global properties of the disks in the high-$z$
ULIRGs appear to be scaled up versions of their local counterparts,
with luminosities $\sim 10 \, \times$ higher and CO-disks $\sim 3 \,
\times$ larger, implying comparable fluxes and $\sds$.  For the $f_g$
and $\sigma$ inferred from the observations, we predict that the
radius at which the disk becomes optically-thick to its own IR
radiation is $R_{\tau_{\rm V}=1}\simeq400-500$ pc.  Our disk
calculations yield a total luminosity of $1-2\times10^{13}$ L$_\odot$
and a flux that varies from $6\times10^{11}$ L$_\odot$ kpc$^{-2}$ at
1.6 kpc to $10^{13}$ L$_\odot$ kpc$^{-2}$ at 200 pc.  We also predict
a dust temperature of $\sim57$ K at $\sim1.6$ kpc, a sound speed of
$c_s\sim60$ km s$^{-1}$, and a gas surface density of
$\Sigma_g\sim0.8$ g cm$^{-2}$.  These predictions are in excellent
agreement with the observations of Tacconi et al.: $L\sim 10^{13} \,
{\rm L_\odot}$, $F\sim 10^{12} \, {\rm L_\odot\,\,kpc^{-2}}$,
$c_s\simeq95\pm30$ km s$^{-1}$ and $\Sigma_g\simeq0.6\pm0.1$ g
cm$^{-2}$.  As discussed extensively above, in local ULIRGs the
nuclear starbursts appear significantly concentrated with respect to
the large-scale CO disks.  It remains to be seen if the same is true
in their high-redshift counterparts (Chapman et al.~2004 find radio
sizes of $\sim 3$ kpc in several systems, suggesting that star
formation may be more extended in some high redshift ULIRGs).


\section{Disk Models with Accretion}
\label{section:disk}

In the previous section we considered the properties of large-scale
starburst disks with constant gas fraction.  In this section we
address the problem of AGN fueling by connecting our kpc-scale starburst disks
with AGN disks on sub-parsec scales.  This requires a consistent
treatment of the gas fraction, which must evolve with radius as a
result of star formation.

On large scales, models with constant gas fraction are equivalent to models
with a constant mass accretion rate.  To see
this, consider a Shakura-Sunyaev accretion disk with viscosity $\nu =
\alpha c_s h$.  The mass accretion rate in such a disk is given by 
\be
\dot M = 2 \pi \nu \Sigma_g \left|\frac{d\ln\Omega}{d\ln r}\right|=
\frac{2^{3/2}\alpha h^3 \Omega^3}{GQ}\left|\frac{d\ln\Omega}{d\ln r}\right|.
\label{mdotvisc}
\ee 
Equating the scale height determined by equation (\ref{mdotvisc})
with that defined in equation (\ref{hzero}) yields a one-to-one
correspondence between the gas fraction and the mass accretion rate:
\be 
f_g=\left(\frac{8}{\alpha Q^2} \left|\frac{d\ln
\Omega}{d\ln r}\right|^{-1}\frac{\dot{M}G}{(\Omega r)^3}\right)^{1/3}.
\label{fgasmdotvisc}
\ee 
Since $\Omega \propto r^{-1}$ in an isothermal potential, equation
(\ref{fgasmdotvisc}) shows that a constant gas fraction implies a
constant accretion rate. On small scales, where the black hole dominates
the gravitational potential, $f_g\propto r^{1/2}$ for constant $\dot{M}$.

In reality, the accretion rate (or $f_g$) is
not constant.  As gas accretes from larger to smaller radii and
some of the gas is converted into stars to maintain $Q\sim1$,
the accretion rate decreases monotonically.  At
the outer radius of the disk $R_{\rm out}$, we assume that gas is
supplied at a rate $\dot{M}_{\rm out}$.  
At any radius $r < R_{\rm out}$ the accretion rate $\dot
M(r)$ is \be \dot{M}(r) =\dot M_{\rm out} - \int^{R_{\rm out}}_r 2 \pi
r\pr \sds \,dr\pr.
\label{mdotevol}
\ee 

If the star formation rate
required to maintain $Q\sim1$ is large enough, all of the gas goes
into stars and the accretion rate becomes vanishingly small.  To
quantify this it is useful to consider two timescales characterizing
the disk: (1) the advection timescale $\tau_{\rm adv} = r/V_r$ and (2)
the star formation timescale $\tau_\star = 1/(\eta \Omega)$.  At any
radius in the disk, there is a critical accretion rate $\dot M_c$ for
which $\tau_{\rm adv} = \tau_\star$.  Taking the optically-thick limit
and assuming $\kappa=\kappa_0T^2$ (\S\ref{section:opacitytemp}) yields
\be \dot{M}_c=r^2\left(\frac{2^{13/2}\pi^3 G Q \sigma_{\rm
SB}}{3\kappa_0^2\epsilon^2c^3}\right)^{1/2}
\hspace{-.4cm}\sim 280 \, r^2_{200}\epsilon_3^{-1}\kappa_{-3.6}^{-1}Q^{1/2}\,{\rm
M_\odot\,yr^{-1}},
\label{mdotc}
\ee
where we have scaled the radius to 200 pc in anticipation of
numerical calculations described below. 
If the gas accretion rate in the disk satisfies $\dot M > \dot M_c$, then
$\tau_{\rm adv} < \tau_\star$ and the gas is able to accrete inwards to
smaller radii.  By contrast if $\dot M < \dot M_c$, then $\tau_{\rm adv} >
\tau_\star$ and most of the gas is converted into stars at radius $r$
without flowing significantly inwards.  We note that gas supply rates
exceeding $\dot M_c$ seem plausible in view of the fact that many
ULIRGs have star formation rates greater than $\dot M_c$.  

Equation (\ref{mdotc}) has several important properties.  First, $\dot
M_c$ is seemingly independent of the viscosity.  This results from the
fact that we have assumed $\kappa \propto T^2$, which is valid only at
low temperatures and thus large radii.  For general $\kappa$ or in
the optically-thin limit,
$\dot{M}_c$ depends explicitly on the viscosity (see Appendix \ref{appendix:mdotcrit}).  
Second, $\dot M_c$
is an {\it increasing} function of radius.  Thus if gas is supplied at
the outer radius $R_{\rm out}$ at a rate $\dot M_{\rm out} > \dot
M_c$, significant gas accretion can continue to smaller radii.
However, on scales smaller than $\sim10$ pc, the central disk
temperature is high enough that dust sublimates, $\kappa$ decreases
dramatically, and the star formation rate required to maintain $Q \sim
1$ increases (Fig.~\ref{plot:disks}; see also Sirko \& Goodman 2003).
The large star formation rate at small radii makes it difficult to
fuel a central accretion disk at a rate sufficient to explain bright
AGN.  This competition between star formation ($\tau_\star$) and
inflow ($\tau_{\rm adv}$), particularly throughout the opacity gap on
$0.1-10$ pc scales, determines the rate at which a central AGN is
fueled.  

\subsection{A Model of AGN Fueling}
\label{section:fueling}

In Appendix \ref{appendix:equations} we collect the equations and
parameters used in this section to derive the properties of starburst
and AGN disks.  As in \S\ref{section:onezone}, we assume that
$\Omega=\Omega_{\rm K}$, but here we account for the gravitational
potential of the black hole: $\Omega_{\rm K}^2=GM_{\rm
BH}/r^3+2\sigma^2/r^2$.  The black hole mass is assumed to be given by
the $M_{\rm BH}-\sigma$ relation: $M_{\rm
BH}\simeq2\times10^8\sigma_{200}^4$ M$_\odot$ (Tremaine et al.~2002;
Ferrarese \& Merritt 2000; Gebhardt et al.~2000).  In calculating
vertical hydrostatic equilibrium, we include gas pressure, which is
important for $r\lesssim1$ pc.  The most important change relative to
the model of \S\ref{section:onezone} is the use of a consistent
accretion rate that accounts for the loss of gas locally to star
formation (eq.~[\ref{mdotevol}]).

Although the critical accretion rate $\dot M_c$ at which $\tau_{\rm
adv} = \tau_\star$ does not depend that strongly on the viscosity in
the disk at large radii (eq.~[\ref{mdotc}]), the fate of gas at small
radii in the opacity gap is a very strong function of the rate of
angular momentum transport.  The efficiency of angular momentum
transport is important because, for fixed $\dot M$, higher viscosity
implies lower surface density and thus self-gravity is comparatively
less problematic.  Indeed, it is well-known that local angular
momentum transport --- such as is produced by the magnetorotational
instability (Balbus 2003) --- is incapable of supplying sufficient gas
to a central black hole to account for luminous AGN (e.g., Shlosman \& Begelman 1989; 
Shlosman et al.~1990; Goodman 2003).  

One possible solution to this problem is that angular momentum
transport proceeds by global torques such as would be provided by
stellar bars, spiral waves, or large-scale magnetic stresses (Shlosman
et al.~1990; Goodman 2003).\footnote{An alternative possibility not
considered here is that AGN are fueled by low angular momentum gas
(Goodman 2003), including perhaps the hot ISM in clusters of galaxies
(e.g., Nulsen \& Fabian 2000).}  In this section, we use a
phenomenological prescription to describe this process: we assume that
the radial transport of gas by a global torque allows the radial
velocity to approach a constant fraction $m$ of the local sound speed
(Goodman 2003).  In this case, \be \dot M = 2 \pi r\Sigma_g
V_r=\frac{2^{3/2}\Omega^3 r h^2 m}{GQ}
\label{mdotdyn}
\ee and the relationship between accretion rate and gas fraction is
given by \be f_g=\left(\frac{2^{3/2}}{Qm}\frac{\dot{M}G}{(\Omega
r)^3}\right)^{1/2}
\label{fgasmdotdyn}
\ee instead of equation~(\ref{fgasmdotvisc}).  Our hope is that, much
as the Shakura-Sunyaev prescription provides a useful zeroth-order
model for local angular momentum transport in disks, the above model
captures some of the essential physics of disks in which angular
momentum transport is dominated by global torques.  With equation
(\ref{mdotdyn}) to relate the gas surface density to the gas accretion
rate, we solve the equations of Appendix \ref{appendix:equations} to determine the
structure of the disk.

\subsection{Results}
\label{section:diskresults}

The dashed lines in Figure \ref{plot:mds} show the mass accretion rate
$\dot{M}$ as a function of radius for $\dot{M}_{\rm out}=80$, 160,
220, 320, and 640 M$_\odot$ yr$^{-1}$ in a model with $\sigma=300$ km
s$^{-1}$ ($M_{\rm BH}\simeq10^9$ M$_\odot$), $R_{\rm out}=200$ pc, and
with dynamical angular momentum transport specified by $m=0.2$.  The
black hole dominates the gravitational potential for $r\lesssim50$
pc. The solid lines show the local star formation rate $\dot{M}_\star$
defined by $\mds=\pi r^2\sds$.\footnote{If $f_g$ exceeds unity at
$R_{\rm out}$, then we expect the disk to become globally
self-gravitating. Through equation (\ref{fgasmdotdyn}), this condition
sets a ``maximum'' mass accretion rate, $\dot{M}_{\rm out}^{\rm
max}\sim1250\sigma_{300}^3 m_{0.2}Q$ M$_\odot$ yr$^{-1}$, above which
the approximations of this paper break down.  For these reasons, we do
not consider models with $\dot{M}_{\rm out}>640$ M$_\odot$ yr$^{-1}$
for the parameters of Figure \ref{plot:mds}.  Equation (\ref{hzero})
also shows that as $f_g\rightarrow1$, $h/r\rightarrow1$.
For models in which $h/r$ or $f_g$ exceeds unity, we expect $m$ or
$\alpha$ to increase because of the large gas fraction (as in the
``bars within bars'' model of Shlosman et al. 1989), a wind to be
driven from the disk surface (thereby limiting $\dot{M}_{\rm out}$
dynamically), or perhaps the gas to be pushed out radially ($R_{\rm
out}$ increases) to maintain $h/r\sim1$.}

There are two types of solution represented in Figure \ref{plot:mds}.
The two models with the smallest $\dot{M}_{\rm out}$ lose
essentially all of the supplied gas at large radii to star formation.
At $r\sim20-40$ pc in these two models the star formation
rate becomes so low that $\dot{M}$ approaches a constant near
$\sim 0.1 \mpy$.  This occurs when the central temperature decreases
sufficiently that gas pressure dominates.
These models have
$\tau_{\rm adv}>\tau_\star$ at $R_{\rm out}$ and $\dot{M}_{\rm
out}<\dot{M}_c(R_{\rm out})$.  They are starburst-dominated and the
star formation occurs predominantly at $\sim R_{\rm out}$.

For the three models shown with $\dot{M}_{\rm out} \ge 220$
M$_\odot$ yr$^{-1}$ the results are qualitatively different.  In these
models, $\dot{M}_{\rm out}$ is large enough that star formation
persists for more than a decade in radius from $R_{\rm out}=200$ pc
down to $r\sim1-10$ pc.  At these small scales, the temperature is
sufficiently hot that dust sublimates and the opacity decreases
sharply (see Figure \ref{plot:kp}).  Because
$\sds\propto\Sigma_g/\kappa$ in the optically-thick limit, the star
formation rate must increase dramatically to maintain $Q\sim1$.  The
increase in $\mds$  as the opacity gap is
encountered is typically an order of magnitude. Although these star
formation rates are large, they are much smaller than those computed
for the constant $\dot{M}$ (or constant $f_g$) solutions in Figure
\ref{plot:disks} (\S\ref{section:onezone}).  In the models with variable $f_g$,
an equilibrium between advection and star formation, expressed by
the equality $\tau_{\rm adv}=\tau_\star$, limits the star formation rate in the
opacity gap (Appendix \ref{appendix:accrete}). Each of
these high $\dot{M}_{\rm out}$ disk models has a gas accretion rate of
$\sim4$ M$_\odot$ yr$^{-1}$ at $r\sim0.01$ pc, sufficient to power a
luminous AGN.  That these models have nearly
identical accretion rates at small radii follows from the fact that $\tau_{\rm adv}=\tau_\star$
in the opacity gap.  
In Appendix \ref{appendix:accrete} we 
present an analytical model for the structure of the disk in this region
and provide simple formulae for the black hole accretion rate.

As Figure \ref{plot:mds} shows, there is a strong bifurcation between
models that fuel a central AGN and those that do not. The critical
mass supply rate distinguishing these two classes of solutions is
close to $\dot M_c$ as estimated in equation (\ref{mdotc}) by
equating $\tau_{\rm adv}$ and $\tau_\star$.\footnote{
As an aside we note that all models presented in Figure \ref{plot:mds} have small $\mds$ at small
radii and one may worry about discreteness; that is, for very small $\mds$
only a few massive stars may be formed in a given annulus and because
an individual massive star only supports a volume $\sim h^3$ around it 
the disk may not be capable of self-regulation.  The ``continuum''
approximation employed throughout this work is only valid when 
$r/h\ll N_{\rm m\star}(r)\sim\mds(r) T_{\rm m\star}10^{-1.35}{\rm M}_\odot^{-1}$,
where $N_{\rm m\star}(r)$ is the number of massive stars at radius $r$,
$T_{\rm m\star}$ is the lifetime of a massive star, and we have assumed a
Salpeter IMF and defined ``massive stars'' as those with mass greater than $\sim10$ M$_\odot$.
For $T_{\rm m\star}$ in the range $10^6-10^7$ yr the continuum approximation is 
valid for the models displayed in Figure \ref{plot:mds}.  For a discussion of when the
continuum limit breaks down, see footnote \ref{footnote:gcdiscrete}.
\label{footnote:discrete}}  

For comparison with these calculations assuming dynamical angular
momentum transport, we have also calculated disk solutions 
employing a local $\alpha$-viscosity
(eq.~\ref{mdotvisc}).  With $\alpha=0.3$, the disk structure exterior
to the opacity gap is very similar to the models with dynamical
angular momentum transport because $h/r$ is large on scales of $R_{\rm out}$
in the models with large $\dot{M}_{\rm out}$.  Therefore, a
local viscosity can transport gas from several hundred parsec scales
down to $\sim10$ pc scales in gas-rich starbursts.  However, we find
that a local viscosity cannot transport gas through the opacity gap at
a rate sufficient to fuel bright AGN;  typical black hole accretion rates 
are several orders of magnitude smaller than those
with dynamical angular momentum transport (see Appendix
\ref{appendix:accrete}).

In the three models with $\dot{M}_{\rm out}\ge220$ M$_\odot$
yr$^{-1} >\dot{M}_c$, we can distinguish between the outer ``starburst disk'',
where the heating of the disk is dominated by star formation
($\sigma_{\rm SB}\teff^4=(1/2)\epsilon \sds c^2$), and the inner ``AGN
disk'', where the heating is dominated by accretion ($\sigma_{\rm
SB}\teff^4=(3/8\pi)\dot{M}(1-\sqrt{R_{\rm in}/r})\Omega^2$). 
The transition between these two
regimes occurs at $R_{\rm AGN}\sim0.5$ pc in the models presented in
Figure \ref{plot:mds}.  At this radius, accretion heating is
sufficient to produce $Q > 1$ and star formation ceases.

By combining the equations describing the ``starburst'' disk at $r >
R_{\rm AGN}$ with those describing the ``AGN disk'' at $r < R_{\rm
AGN}$ we have computed the full radial structure of models from
hundreds of parsecs down to the central black hole.   
We note again that there are multiple solutions at some radii and that
we have selected what we believe is the physical disk solution as
described in Appendix \ref{appendix:stability}.  As an example of the
full radial disk structure, Figure \ref{plot:allp} shows the results
for the model with $\dot{M}_{\rm out}=320$ M$_\odot$ yr$^{-1}$ (see
also Fig.~\ref{plot:mds}).  In the middle left panel, note that gas
pressure dominates radiation pressure over $\sim1$ decade in radius
inside the opacity gap at $r\lesssim1$ pc (Appendix
\ref{appendix:accrete}), but that radiation pressure dominates at all
other radii.\footnote{ In these calculations, we have assumed that
dynamical angular momentum transport acts throughout the disk.  On
sub-parsec scales near the black hole, perhaps at $R_{\rm AGN}$ where
$Q$ becomes greater than unity, we expect accretion to be driven
instead by local viscosity. This will modify the density, scale
height, etc., close to the black hole ($\lesssim 0.5$ pc in Figure
\ref{plot:allp}), but will not change our conclusions about the need
for angular momentum transport by global torques at larger radii.}

Assuming that every annulus in the disk radiates as a blackbody and
that the disk is viewed face-on, the spectral energy distribution of
our disk models can be computed using \be \lambda
L_\lambda =\frac{2\pi hc^2}{\lambda^4}\int_{R_{\rm in}}^{R_{\rm
out}}\frac{2\pi r dr}{\exp[hc/\lambda k_{\rm B}\teff(r)]-1}.
\label{sed}
\ee The resulting multi-color blackbody spectra for the models
presented in Figure \ref{plot:mds} are shown in Figure
\ref{plot:spect}, assuming an inner disk radius of $R_{\rm
in}=3(2GM_{\rm BH}/c^2)$.  The models with small mass supply rate
($\dot{M}_{\rm out}<\dot{M}_c$) have the lowest bolometric luminosity
and virtually all of their flux is generated by the starburst at
$R_{\rm out}$.  The spectra of these models peak at $\sim50$ $\mu$m,
corresponding to an outer disk dust temperature of $\sim70$ K. For
higher $\dot{M}_{\rm out}$, the dust temperature at $R_{\rm out}$
increases somewhat and the spectral peak moves to shorter wavelengths.

Higher $\dot M_{\rm out}$ models have similar FIR spectra.  However,
at $\sim8$ $\mu$m they exhibit an additional spectral peak.
This peak is produced by the inner ring of star formation on $1-10$ pc
scales shown in Figure \ref{plot:mds}, which is caused by the decrease
in the opacity of the disk at the dust sublimation temperature,
$\tsub\sim1000$ K.  Although $\tsub$ sets the central temperature of
the disk where this sudden increase in $\sds$ occurs, the disk is
typically optically thick at these radii and the effective temperature
can be a factor of $\sim 3-4$ smaller, accounting for the fact that
the spectral peak associated with star formation in the opacity gap is
at $\sim8$ $\mu$m.
Together, the FIR peak from the
outer starburst ring and the peak from the inner starburst ring
create a broad IR peak spanning $\sim1.5$ decades in wavelength.  

The distinguishing feature of the models with large $\dot M_{\rm out}$
is the significant contribution of the AGN to the bolometric
luminosity: the UV emission in Figure \ref{plot:spect} is comparable
to or dominates the peaks at $\sim8$ and $\sim50$ $\mu$m.  Figure
\ref{plot:spect} suggests that quasar accretion disks can plausibly be
fed by radiation pressure supported starburst disks at larger radii.
Indeed, the broad but sub-dominant IR peak and the magnitude of the
AGN luminosity in our models are reasonably consistent with composite
quasar spectra (Elvis et al.~1994; Haas et al.~2003).  

It should be noted that the spectra shown in Figure \ref{plot:spect}
are photospheric and do not include AGN reprocessing. Because the disk
may intercept some of the AGN emission, we expect that reprocessing
could lead to an additional emission component at $\sim 1-2 \mu$m if
dust in the surface layers of the disk is heated to the sublimation
temperature.

\subsection{Vertical Structure \& Nuclear Obscuration in AGN}
\label{section:torus}

At radii $\sim 1$ pc the models shown in Figure \ref{plot:allp} have
the interesting property that the central temperature of the disk is
above the sublimation temperature of dust while the effective
(surface) temperature of the disk is below the sublimation
temperature.  As a result, there is a large {vertical} gradient in the
opacity of the disk.  Because the disk is radiation pressure
dominated, this large gradient in the opacity has important
implications for the disk's vertical structure.  Here we argue that
the photosphere of the disk can lie a distance $h_{ph} \sim r$ off of
the midplane even though the midplane scale height is small, with $h
\ll r$ (where $h = c_s/\Omega$ is evaluated using the midplane
properties of the disk).  We then discuss the implications of this
result.

The equations governing the vertical structure of the disk include
those of vertical hydrostatic equilibrium and vertical energy
transport: \be {dp \over dz} = - \rho \Omega^2 z,
\label{he} \ee where $p = p_{\rm gas} + p_{\rm rad}$ and we assume
that $\Omega$ is independent of the height $z$ in the disk.  Assuming
that energy is transported diffusively by photons implies \be {dT
\over dz} = - {3 \kappa \rho F \over 16 T^3 \sigma_{\rm SB}}
\label{diff} \ee where $F$ is the
vertical flux of energy.  If the disk is radiation pressure dominated,
as our models are at most radii, then equations (\ref{he}) and
(\ref{diff}) imply that the flux in the disk must be everywhere equal
to the Eddington flux: \be F = F_{\rm EDD} = {\Omega^2 c z \over
\kappa}. \label{edd} \ee

The presence of a large vertical opacity gradient suggests that
convection might develop in the outer atmosphere of the disk.
However, two arguments show that convection is unlikely to be
important in the present context: (1) Our disk models have $\tau_{\rm th}
\Omega \ll 1$, where $\tau_{\rm th}$ is the cooling time of the disk
(Appendix \ref{appendix:stability}).  This implies that the diffusion
time through the disk is much shorter than the characteristic
timescale of buoyant convective motions. In the presence of such
strong radiative diffusion, convective modes driven by radiation
entropy gradients grow very slowly, if at all (e.g., Blaes \& Socrates
2003).  (2) Even if convection develops, it is unlikely to carry a
significant fraction of the vertical energy flux.  Using $F \approx
\rho V_c^3$ to estimate the convective velocity $V_c$ required to
carry the energy flux we find very high Mach numbers: 
\be {V_c \over
c_s} \approx 40 \, M_9^{1/3} \left(z \over r\right)^{1/3} \left(F
\over F_{\rm EDD}\right)^{1/3} r_{\rm pc}^{-2/3} T_{1000}^{-2}
\kappa_1^{-1/3} \rho_{-15}^{1/6}, \label{vc} \ee where $\rho_{-15}=\rho/10^{-15}$ g cm$^{-3}$ and
we have scaled our estimate to parameters appropriate to parsec scale-disks (see
Fig.~\ref{plot:allp}).  These considerations strongly suggest
that radiative diffusion dominates the vertical transport of energy
throughout the disk.

Unfortunately, equations (\ref{he}) and (\ref{diff}) are insufficient
to fully specify the vertical structure of the disk.  An equation for
the vertical generation of energy, $dF/dz$, is also needed (e.g.,
Davis et al. 2005).  The vertical heating profile is particularly
uncertain in the present context where star formation, rather than
accretion, dominates the heating of the disk.  Given this
significant uncertainty, we restrict ourselves to the following simple
considerations about the disk's vertical structure.  

We assume that the calculations described in the previous sections ---
that apply to the midplane of the disk --- provide an adequate estimate
of the star formation rate $\dot \Sigma_\star$ and, thereby, the flux $F$ required
to maintain $Q \sim 1$.  This assumes that the atmosphere of the disk
does not contribute much flux, a reasonable
assumption.  Because the disk is radiation pressure dominated near its
photosphere, we can estimate the location of the photosphere using $F
\approx F_{\rm EDD}$.  This yields \be h_{ph} \approx {F
\kappa_{ph} \over \Omega^2 c} \approx h \left({\kappa_{ph}
\over \kappa_{\rm mid}}\right),
\label{photo} \ee where $\kappa_{ph}$ is the opacity at the
photosphere, evaluated using the effective temperature, and
$\kappa_{\rm mid}$ is the opacity at the midplane, evaluated
using the central temperature.  Equation (\ref{photo}) shows that if
the opacity at the surface of the disk is much larger than the opacity
near the midplane, as will inevitably occur in the vicinity of the
sublimation temperature of dust, then $h_{ph} \gg h$ and the
midplane scale height does not provide a good estimate of the location
of the photosphere of the disk.

Figure \ref{plot:hr} shows our estimates of the midplane scale-height
$h/r$ and the photospheric scale-height $h_{ph}/r$ for models with
$\dot{M}_{\rm out}=320$ and 640 M$_\odot$ yr$^{-1}$ (whose other
properties have been described in Figures
\ref{plot:mds}-\ref{plot:spect}).  These estimates of the photospheric
scale-height show that $h_{ph}$ can readily approach  $\sim r$ at
$\sim 0.1-10$ pc, even though the midplane scale-height of the disk
satisfies $h \ll r$.  
Because the effective temperature of the disk is
$\lesssim \tsub$, the opacity near the photosphere is $\kappa \sim 1$
g cm$^{-2}$ and the hydrogen column through the disk's atmosphere can
approach $\sim 10^{24}$ cm$^2$.  
We suggest that this extended dusty atmosphere may account for the
presence of obscuring material inferred from observations of AGN
(e.g., Antonucci 1993).  In particular, although AGN disks are
typically classified as being ``thin'' because $h \ll r$, Figure
\ref{plot:hr} shows that the photosphere may nonetheless puff up
substantially and reach $h_{ph} \sim r$.  Some of this material is
likely to be unbound and the origin of a dusty outflow.

The disk's extended atmosphere is supported by radiation pressure from
a parsec-scale nuclear starburst (in contrast to Pier \& Krolik
1992, who argued that dusty gas could be supported at $h \sim r$ by
radiation pressure from the AGN).  Estimating the luminosity of the disk
at these radii, $L \equiv \pi r^2 F$, we find \be L \approx {\pi G M c
\over \kappa_{ph}} \left(h_{ph} \over r \right) \approx
{L_{\rm es} \over 4} \left({\kappa_{\rm es} \over \kappa_{\rm
ph}}\right) \left(h_{ph} \over r \right),
\label{ltorus} \ee where $L_{\rm es}$ is the canonical Eddington 
luminosity defined using the electron-scattering opacity $\kappa_{\rm
es}$.  Taking $\kappa_{ph} \sim 10 \, \kappa_{\rm es}$, equation
(\ref{ltorus}) implies that in order to have $h_{ph} \sim r$, as
suggested by the relative number of Type 1 \& Type 2 Seyfert galaxies,
the luminosity of the parsec-scale disk must be $\sim 0.01-0.1 \,
L_{\rm es}$.  Therefore, in order for the disk to have $h_{ph}\sim r$, we predict 
the presence of a very compact nuclear starburst with a luminosity similar 
to that of the AGN.

Because the luminosity from star formation required to ``puff up'' the
disk is substantial, it is unlikely that this model is relevant to
low-luminosity Seyfert 2 galaxies such as NGC 4258 (where there is
also strong evidence that the nuclear obscuration is due to a warped
disk; e.g., Fruscione et al.~2005).  A compact starburst may, however,
dynamically support obscuring material in more luminous AGN.  In
particular, there is evidence for a compact starburst in the
prototypical Seyfert 2 Galaxy NGC 1068.  IR
interferometry of 1068 has also directly resolved warm dusty gas on
the parsec scales predicted by our model (Jaffe et al.~2004; Rouan et al.~2004).


\section{Discussion \& Conclusions}
\label{section:discussion}

In standard models of galactic-scale star formation, energy and
momentum injected by supernovae and stellar winds are assumed to drive
turbulent motions in the interstellar medium of a galaxy (e.g., Silk
1997).  This turbulent pressure helps stave off the self-gravity of
the disk and maintain marginal stability to gravitational
perturbations (Toomre's $Q \sim 1$).  In this paper, we have focused
instead on the role of radiation pressure on dust grains in regulating
the structure and dynamics of star formation in galaxies.  To order of
magnitude, the turbulent pressure from supernovae, stellar winds, and
the radiation from massive stars are all comparably important when the
galactic disk is optically thin to its own IR radiation, as in normal
star-forming galaxies.  By contrast, when the disk is sufficiently
optically thick to the IR, radiation pressure provides the dominant
vertical support against gravity.  This condition is met in the inner
few hundred parsecs of luminous gas-rich starbursts, most notably in
Ultraluminous Infrared Galaxies (ULIRGs).  In addition, the outer
parts of accretion disks around AGN are expected to be dominated by
radiation pressure on dust.  Understanding the dynamics of disks in
the radiation pressure dominated limit is therefore crucial for
understanding nuclear starbursts and AGN fueling.

We have constructed simple quantitative models of self-regulated disks
appropriate to both the optically-thin and optically-thick limits.
From these models, we derive the star formation rate per unit area
required to maintain $Q\sim1$.  In the optically-thin limit, we find
that $\sds\propto\Sigma_g^2$ (eq. [\ref{sds1scat}]), a slightly steep
version of the Schmidt law for star formation, whereas in the
optically-thick limit, $\sds\propto\Sigma_g^2/\tau_V \propto
\Sigma_g/\kappa$ (eq. [\ref{sds}]).  Because radiation pressure
dominates the vertical pressure support in the optically-thick limit,
the star formation rate is sensitive to the mean opacity of the disk, 
and thus also to the metallicity. 
In fact, each annulus of the disk radiates at its local
Eddington limit, defined using the Rosseland mean opacity $\kappa$.  
This criterion can be written as (see also Scoville
2003) $L/M\sim10^3\fgas\kappa_1^{-1}$ L$_\odot$/M$_\odot$, where the
gas fraction $f_g$ is proportional to the disk thickness $h/r$.  

Our prediction that the Schmidt law changes in the optically-thick
limit (with $\sds\propto\Sigma_g/\kappa$) is currently difficult to
test and requires further high resolution observations.  It is
important to stress that in our models the star formation efficiency
in the optically-thick nuclei of starbursts is actually {\it lower}
than what would be inferred from an extrapolation of the star
formation efficiency in normal star-forming galaxies because of the
extra support from radiation pressure in the optically-thick limit.

Our calculations show that in the optically-thick portion of starburst
disks near $\sim 100$ pc, the flux, star formation rate per unit area, and
the effective temperature of the disk are roughly constant with
characteristic values of $F\sim10^{13}$ L$_\odot$ kpc$^{-2}$, 
$\dot \Sigma_\star \sim 10^3 \mpykpc$, and
$T_{\rm eff} \sim 90$ K (\S\ref{section:opacitytemp}). To test these
predictions, we have estimated the fluxes (temperatures) in the
emission regions of ULIRGs using the resolved radio images of Condon
et al.~(1991) as indicative of the sizes of nuclear starbursts in
these systems (the fluxes inferred using FIR blackbody temperatures may not accurately
characterize the fluxes in the emission region if there is still
significant obscuration at $\gtrsim20-30$ microns).  
We find excellent agreement with the predictions of
our models (Figs. \ref{plot:fluxdist} \& \ref{plot:fluxr}).  This
supports our interpretation that a significant fraction of the
radiation from ULIRGs is produced by an Eddington-limited starburst.
In the future, imaging at optically-thin FIR wavelengths will test our
assumption that the radio emission traces star formation in ULIRGs.

In our model of starburst disks, the central temperature exceeds the
dust sublimation temperature ($T_{\rm sub} \sim 1000$ K) at radii
$\sim 1-10$ pc.  The opacity of the disk then drops dramatically and
the star formation rate required to maintain $Q\sim1$ is very large
(Fig.~\ref{plot:disks}).  This poses a severe problem for fueling a
central AGN with gas stored in the starburst disk at larger radii
(Sirko \& Goodman 2003).  At first glance it appears that all of the
gas is converted into stars in the ``opacity gap'' with little left 
available to power a central AGN.

A number of solutions to this problem have been proposed, all of which
hinge on the efficiency of angular momentum transport in the disk
(Shlosman \& Begelman 1989; Goodman 2003).  For fixed $\dot M$, higher
viscosity implies lower surface density and thus self-gravity is
comparatively less problematic.  For this reason, we have focused on
the possibility that angular momentum transport proceeds via a global
torque, as would be provided by a bar, spiral waves, or a large-scale
wind.
With this prescription for angular momentum transport we have extended
our starburst disk models to smaller radii where they connect
consistently with AGN disks on sub-parsec scales.  A key ingredient to
these calculations is an equation that accounts for the radial loss of
gas due to star formation (eq. [\ref{mdotevol}]).   
This limits the star formation rate  required to support the disk in the opacity gap region 
and, therefore,  allows for significant gas accretion and AGN fueling
(Appendix \ref{appendix:accrete}). In this respect our
model differs from that of Sirko \& Goodman (2003), who also considered
star formation-supported AGN disks, but assumed constant $\dot{M}$.

Our model provides a way of quantifying 
the connection between starbursts and AGN activity.  In particular,
we find two classes of disk models (Figs.~\ref{plot:mds}-\ref{plot:spect}): 
(1) For mass supply rates at the outer edge of the
disk less than a critical rate $\dot{M}_c$ (eq.~[\ref{mdotc}]), a
starburst occurs primarily in a narrow ring at $R_{\rm out}$.  Nearly
all of the gas is converted into stars at large radii and the spectrum
is dominated by a FIR starburst peak.  This class of
solutions corresponds to the limit in which the star formation time in
the disk is shorter than the viscous time ($\tau_\star<\tau_{\rm adv}$).  
(2) On the other hand, for
$\dot{M}_{\rm out}>\dot{M}_c$, a fraction of the gas accretes
inwards to smaller radii because the advection timescale is shorter than the
star formation timescale.  These disk solutions have an outer starburst at $\sim R_{\rm out}$ 
and an inner nuclear starburst ring on $\sim1-10$ pc scales at the
opacity gap (Fig.~\ref{plot:mds}).  In addition, for fiducial parameters
we find that the accretion rate onto the central BH is $\sim 1-10$
M$_\odot$ yr$^{-1}$ (Appendix \ref{appendix:accrete}).  Models with $\dot{M}_{\rm out}>\dot{M}_c$
can be dominated by AGN emission, although there is a prominent
contribution in the mid- and far-IR from the inner and outer starbursts
(Fig.~\ref{plot:spect}).  Both the broad but sub-dominant IR peak and
the magnitude of the AGN luminosity in our models are reasonably
consistent with composite AGN spectra.

One prediction of our models is that a significant fraction of the IR
emission in luminous AGN must be from star formation rather than
reprocessing.  The star formation is required to support the accretion
disk at large radii against its own self-gravity (see Rowan-Robinson
2000 for an observational discussion of this point).  We have
identified an additional observational consequence of the parsec-scale
starburst that is predicted to occur when the temperature of the disk
reaches the sublimation temperature of dust.  Our estimates suggest
that for luminous AGN, the nuclear starburst is able to inflate the
photosphere of the disk to a height $h_{ph} \sim r$ (\S
\ref{section:torus} and Fig. \ref{plot:hr}).  This extended atmosphere
of the disk contains little mass but produces significant obscuration,
and may account for some of the nuclear obscuration observed in Type 2
AGN.

As a final application of our calculations, we note that there is
direct observational evidence for compact starbursts near massive
black holes.  In particular, observations of the Galactic Center
reveal a population of young O and B stars within the central parsec
of our Galaxy.  Most of these stars appear to lie in a thin disk
located within $0.1-0.3$ parsecs of the black hole (Levin \&
Beloborodov 2003; Genzel et al. 2003).  The origin of these stars
remains uncertain, but their disk-like kinematics suggests that they
might have formed in a dense self-gravitating accretion disk, if the
black hole in the Galactic Center were accreting at a much higher rate
several million years ago (e.g., Levin \& Beloborodov 2003).  To
explore this hypothesis in the context of our models, Figure
\ref{plot:gc} shows the star formation rate and accretion rate as a
function of radius in models with $\sigma = 75 \kms$, $M_{\rm BH} = 4
\times 10^6 M_\odot$, $R_{\rm out} = 3$ pc, and an $\alpha$-viscosity
with $\alpha = 0.3$ (we also increased the opacity by a factor of
three relative to the curves presented in Figure \ref{plot:kp} to
account for the super-solar metallicity of the Galactic Center).  The
chosen outer radius is appropriate if the circum-nuclear disk in the
Galactic Center --- a current reservoir of $\sim 10^4-10^5 \,M_\odot$
of gas (e.g., Jackson et al.~1993; Shukla, Yun, \& Scoville 2004) ---
were accreted onto the central black hole.  The mass supply rates we
have used --- $\dot M_{\rm out} = 0.015-0.15 \mpy$ --- are in the
range expected if the circum-nuclear disk were accreted on a viscous
timescale $\sim 10^6$ years.  Figure \ref{plot:gc} shows that the star
formation rate in this hypothesized burst of accretion typically has a
strong peak at $r \sim 0.1$ pc.  This is the ``inner nuclear
starburst'' caused by the decrease in opacity when the central disk
temperature exceeds the sublimation temperature of dust.  The location
of this starburst is strikingly similar to the current location of the
stellar disk in the Galactic Center.\footnote{
For the models in Figure \ref{plot:gc} that produce a peak in $\mds$ at $r\sim0.1$ pc,
the continuum approximation (see footnote \ref{footnote:discrete}) breaks
down for $r\lesssim0.05$ pc.  In this region the disk will not self-regulate and
it will cool rapidly and form more stars.  The star formation rate in this
region will then be determined by the criterion $r/h\sim N_{\rm m\star}$
and not $Q\sim1$.
Because the continuum limit is violated by a factor of $\sim3-5$ in the models
displayed in Figure \ref{plot:gc} for $r\lesssim0.05$ pc, $\mds$ 
will be $\sim3-5$ times higher in this region. The gas accretion rate 
at yet smaller radii will be only somewhat smaller. \label{footnote:gcdiscrete}
}

Several aspects of our calculations require further study.  The first
is the stability of our disk solutions.  It is well known that
radiation-pressure supported disks are prone to numerous
instabilities: viscous, thermal, convective, and photon-bubble (e.g.,
Lightman \& Eardley 1974; Piran 1978; Blaes \& Socrates 2003).  Such
instabilities have been extensively studied in the context of the
central $\sim 10-100$ Schwarzschild radii of black hole accretion
disks, where the radiation pressure is on free electrons. An important
difference between our solutions and these is that at large radii the
disk is heated primarily by star formation, not accretion.  We find
that as a result, the disk is globally thermally and viscously stable
(Appendix \ref{appendix:stability}).  As in the local ISM, however,
the disk may fragment into a multi-phase medium.  A second aspect of
our model that requires further study is the physics of star formation
under the unusual conditions appropriate to gravitationally unstable
accretion disks, particularly at radii $\sim$ pc where the gas is much
denser and hotter than in normal star-forming regions.  Because $Q \ll
1$ in the absence of star formation, and because the cooling time of
the disk is much shorter than the orbital time (Appendix
\ref{appendix:stability}), we believe that fragmentation of the disk
is inevitable.  But beyond this, the properties of star formation
under these conditions are poorly understood (Goodman \& Tan 2004
argue that the disk will form supermassive stars).  
Interestingly, the power-law solutions for the star formation rate 
in the opacity gap region (Appendix \ref{appendix:accrete}) 
are independent of the IMF-dependent parameter $\epsilon$ (eq.~\ref{prp}).
On larger scales ($r\gtrsim1-10$ pc), however, a top-heavy IMF (larger $\epsilon$) 
can suppress the star formation rate (eqs.~\ref{sds1scat} \& \ref{sds}), decrease
the star formation efficiency $\eta$ (eqs.~\ref{eta1scat} \& \ref{eta}), and 
decrease the critical mass accretion rate required for AGN fueling (eq.~\ref{mdotc}).
Another component
of our model requiring further study is the interaction of the AGN
emission with the flared starburst disk and the tenuous dusty
obscuring atmosphere predicted in \S\ref{section:torus}.

Finally, we note that large-scale outflows of cold dusty gas are
commonly observed from starbursting galaxies, including ULIRGs (Alton
et al.~1999; Heckman et al.~2000; Martin 2004).
We have recently proposed that radiation pressure on dust is an
important mechanism in driving such winds (Murray, Quataert, \&
Thompson 2005 [MQT]).  The disk models presented here provide support
for this idea because it is natural to suspect that a radiation
pressure supported starburst disk will drive a wind from its
photosphere, much as near-Eddington stars drive strong outflows.  In
MQT we further argued that both starbursts and AGN have a maximum
Eddington-like luminosity given by $L_{\rm
M}=4f_g\sigma^4c/G$.\footnote{The ``M'' in $L_{\rm M}$ stands for
``momentum-driven''.}  One can show that the disk models presented in
this paper have luminosities $\lesssim L_M$, with the equality
obtained (approximately) when $h \sim r$ and the spherical limit
employed in MQT is realized.  We also note that the results in this
paper, together with those of MQT, imply that fueling a luminous AGN
requires an accretion rate in the range $\dot{M}_c \lesssim
\dot{M}_{\rm out} \lesssim \dot{M}_{\rm max} \equiv L_M/(\epsilon
c^2)$.  With the black hole accretion rate given by equation
(\ref{mdotbh}), this yields a ratio of star formation rate to black
hole accretion rate of $\sim 100-1000$, similar to the observed ratio
of stellar mass to BH mass in nearby galaxies (Magorrian et al.~1998;
H\"{a}ring \& Rix 2004).


\acknowledgments We thank Reinhard Genzel and Chris McKee for a number
of stimulating conversations and for a critical reading of the text.
We also thank Omer Blaes, Shane Davis, Bruce Draine, Jeremy Goodman,
James Graham, Yoram Lithwick, and Linda Taconni for useful
discussions, and Semenov et al.~for making their dust opacities
publicly available.  T.A.T. is supported by NASA through Hubble
Fellowship grant \#HST-HF-01157.01-A awarded by the Space Telescope
Science Institute, which is operated by the Association of
Universities for Research in Astronomy, Inc., for NASA, under contract
NAS 5-26555.  E.Q. is supported in part by NSF grant AST 0206006, NASA
grant NAG5-12043, an Alfred P. Sloan Fellowship, and the David and
Lucile Packard Foundation.  N.M. thanks the Miller Foundation for
supporting his stay at UC Berkeley.  N.M. is also supported in part by
a Canadian Research Chair in Astrophysics.



\newpage

\begin{appendix}

\section{Appendix A: The Black Hole Accretion Rate}
\label{appendix:accrete}

An interesting feature of the results shown in Figure \ref{plot:mds}
is that all of the models with $\dot{M}_{\rm out}>\dot{M}_c$ deliver
precisely the same amount of mass to the AGN disk at small radii.
More specifically, all of the dashed lines for $\dot{M}_{\rm
out}\ge220$ M$_\odot$ yr$^{-1}$ in Figure \ref{plot:mds} come together
and attach to a single power-law solution in the opacity gap between
$r\sim1-10$ pc, and then settle onto a constant $\dot M \approx 4
\mpy$ solution at smaller radii.  In this appendix we explain why this
is the case.

The steep drop in the opacity in Figure \ref{plot:kp} occurs when the
central disk temperature reaches the sublimation temperature $\tsub$
at a radius $R_{\rm sub}$.  In the models presented in Figure
\ref{plot:mds} with $\dot{M}>\dot{M}_c$, $R_{\rm sub}$ corresponds to
the sudden increase in $\mds$ between $\sim1$ and $\sim10$ pc.  The
rapid decrease in $\kappa$ inside the opacity gap can be approximated
by a power law, $\kappa(T\gtrsim\tsub)=\ksub T^{\beta}$.  For example,
in the opacity models of Bell \& Lin (1994), $\beta=-24$ and
$\tsub\sim900$ K (see Fig.~\ref{plot:kp}).  Because the star formation
rate required to maintain $Q \sim 1$ in the optically-thick limit is
given by $\sds\propto\Sigma_g/\kappa$ (eq. [\ref{sds}]), one might
expect $\sds$ to increase by $\sim5$ orders of magnitude inside the
opacity gap because $\kappa$ decreases by this amount at $T_{\rm sub}$
(see Fig.~\ref{plot:kp}).  In the models with constant $f_g$ shown in
Figure \ref{plot:disks}, this is precisely what happens.  However,
when we self-consistently solve for $\dot M(r)$
(eq.~[\ref{mdotevol}]), as in the models presented in Figure
\ref{plot:mds}, the results are qualitatively different.  The star
formation rate increases by only $\sim1$ order of magnitude at $r\sim
R_{\rm sub}$ because of an important physical effect.  As $\kappa$
decreases and $\mds$ increases, the accretion rate must decrease
because the gas is depleted as a result of star formation.  This leads
to a lower surface density and, therefore, a lower star formation rate
is required to maintain $Q\sim1$.  Thus, an equilibrium is established
between star formation and advection in the opacity gap ($\tau_{\rm
adv}=\tau_\star$).  An analytic solution representing this balance
between star formation and accretion can be derived by solving
equation (\ref{mdotevol}) for $\dot M(r)$ using $\kappa \propto
T^{\beta}$, $T\propto\Sigma_g^{1/2}$ (eq. [\ref{temp}]),
$\sds\propto\Sigma_g/\kappa$ (eq. [\ref{sds}]), and the fact that there
is a one-to-one relationship between $f_g$ and $\dot M$.  If angular
momentum transport proceeds via global torques, the $f_g-\dot M$
relation is given by equation (\ref{fgasmdotdyn}), and the analytic
solution for $\dot M(r)$ takes the form \beqa \dot M(r)^{1/2 +
\beta/4}- \dot M(\rsub)^{1/2 + \beta/4} = A \left(r^{3/4 + 5 \beta/8}
- \rsub^{3/4 + 5 \beta/8}\right),
\label{mdotopac}
\eeqa where $A$ is a $\beta$-dependent  constant that will be specified below.  Because
$\beta$ is $\ll 0$ in the opacity gap, the solution quickly loses
memory of the initial conditions at $\rsub$.  As a result, a single
universal equilibrium power law solution for $\dot{M}\approx\mds$ exists in the opacity gap.
Because of this equilibrium between star formation and advection, the increase in $\mds$
at $R_{\rm sub}$ is much
less pronounced in these solutions than in the constant $f_g$ models
(compare Fig. \ref{plot:disks} \& \ref{plot:mds}).

The general expression for the constant $A$ for arbitrary $\beta$ is a
bit awkward, but the limit $\beta\rightarrow-\infty$ (as appropriate for a step function in opacity) 
provides a useful approximation.  In this limit, the accretion rate inside the opacity
gap is given by \be \dot M(r)\approx \frac{4\sigma_{\rm
SB}}{3c}\tsub^4\,\frac{4m\pi r}{\Omega}
\approx 720 \, T_{1000}^4m_{0.1}r_{10}^{5/2}M_9^{-1/2}\,{\rm
M_\odot\,yr^{-1}},
\label{mdoteq}
\ee where $T_{1000}=\tsub/1000$K, $m_{0.1}=m/0.1$, $r_{10}=r/10$pc,
and $M_9=M_{\rm BH}/10^9$M$_\odot$.  Note that the normalization of
$\dot M$ in equation (\ref{mdoteq}) depends only on the black hole
mass $M_{\rm BH}$, the Mach number $m=V_r/c_s$, and the dust opacity law through
$\tsub$.  This explains why all of the solutions with $\dot{M}_{\rm
out}>\dot{M}_c$ in Figure \ref{plot:mds} are the same inside the
opacity gap.  Note also that because $\dot M(r)$ is a power-law inside
the opacity gap, the star formation rate $\mds(r)$ is as well (see
eq. [\ref{mdotevol}]).  Indeed, for these solutions, $\mds(r)
\equiv \pi r^2 \sds = (5/4) \dot M(r)$, in good agreement with Figure
\ref{plot:mds}.  Finally, because $\dot{M}_\star\propto r^{5/2}$ in
the opacity gap, both $\teff^4$ and $\sds$ are proportional to
$r^{1/2}$.  The central temperature $T$ and the surface density
$\Sigma_g$, on the other hand, are independent of radius.

As is clear from Figure \ref{plot:mds}, the equilibrium power law
solution for $\dot{M}$ given in equation (\ref{mdoteq}) does not
persist to arbitrarily small radii.  Because $\rho$ increases and $T$
is constant in the opacity gap, gas pressure eventually dominates
radiation pressure.  The two pressures are comparable at a radius
$R_{\rm EOS}$ given by \be R_{\rm EOS}\sim 1 \,
M_9^{5/9}\dot{M}_1^{-2/3}m_{0.1}^{2/3}Q^{-8/9}\,{\rm pc},
\label{reos}
\ee where $\dot{M}_1=\dot{M}/1$ M$_\odot$ yr$^{-1}$.  For $r<R_{\rm
EOS}$, one can show that star formation is no longer a significant
impediment to accretion: $\mds$ drops precipitously and $\dot{M}$
approaches a constant.  As a result, we can estimate the accretion
rate onto the black hole by evaluating equation (\ref{mdoteq}) at $r
\sim R_{\rm EOS}$.  This yields 
\be 
\dot{M}_{\rm BH}=2^{9/4}\tsub^{3/2}m M_{\rm
BH}^{1/3}G^{-1/2}\left(\frac{\sigma_{\rm SB}\pi}{3c}\right)^{1/6}
\left(\frac{k_{\rm B}}{Qm_p}\right)^{5/6} \sim 2\,
T_{1000}^{3/2}m_{0.1}M_9^{1/3}Q^{-5/6} \,{\rm M_\odot \,\,yr^{-1}},
\label{mdotbh} 
\ee where we emphasize that $m$ is the Mach number in the opacity gap
region; even if angular momentum transport becomes inefficient for
$r<R_{\rm EOS}$ (e.g.~$m\ll0.1$ or a transition to local viscosity
with $\alpha\lesssim0.1$ occurs), $\dot{M}_{\rm BH}$ remains
unchanged.  Using equation (\ref{mdotbh}) and taking $L_{\rm BH}=\zeta
\dot{M}_{\rm BH}c^2$ we find that \be L_{\rm
BH}\sim10^{46}\zeta_{0.1}T_{1000}^{3/2}m_{0.1}M_9^{1/3}Q^{-5/6} \,{\rm
ergs\,\,s^{-1}}, \ee where $\zeta_{0.1}=\zeta/0.1$.  This result is in
excellent agreement with the models presented in Figures
\ref{plot:mds} and \ref{plot:spect}.  With $L_{\rm EDD}=\zeta
\dot{M}_{\rm EDD}c^2=4\pi GM_{\rm BH}c/\kappa_{\rm es}$, where
$\kappa_{\rm es}$ is the electron scattering opacity, we find that
$\dot{M}_{\rm BH}/\dot{M}_{\rm EDD}\sim0.1
m_{0.1}\zeta_{0.1}M_9^{-2/3}$.  Therefore, for typical parameters, the
black hole is fed at a reasonable fraction of its Eddington rate.
Because of the scaling with $M_{\rm BH}$ we expect black holes of
smaller mass to be preferentially super-Eddington.

Note again that from equation (\ref{mdotbh}), the accretion rate
through the opacity gap into the AGN disk depends primarily on the
dust sublimation temperature and on the rate of angular momentum
transport in the opacity gap (via the Mach number $m$).  Because of
the latter dependence, it is of interest to compare the above
expressions with the analogous results assuming that angular momentum
transport in the opacity gap is via a local viscosity, for which the
$f_g-\dot M$ relation is given by equation (\ref{fgasmdotvisc}).  In
this case the accretion rate in the opacity gap is given by
\be 
\dot M(r)=\frac{3\sqrt{2}\alpha}{GQ}\left(\frac{4\sigma_{\rm SB}}{3c}T_{\rm sub}^4\sqrt{2}\pi G Q\right)^{3/2}
\frac{1}{\Omega^3}
\approx 210 \alpha_{0.1}T_{1000}^6Q^{1/2}M_9^{-3/2}r_{10}^{9/2}\,{\rm
M_\odot\,yr^{-1}},
\label{mdoteqvisc}
\ee 
and the accretion rate onto the black hole is given by
\be 
\dot{M}_{\rm BH}=\frac{3\sqrt{2}\alpha}{GQ}\left(\frac{k_{\rm B}T_{\rm sub}}{m_p}\right)^{3/2}
\approx2\times10^{-3}\alpha_{0.1}Q^{-1}T_{1000}^{3/2}\,{\rm M_\odot \,\,yr^{-1}}.
\label{mdotbhvisc} 
\ee Equations (\ref{mdoteqvisc}) and (\ref{mdotbhvisc}) highlight the
fact that in our models, local viscosity is insufficient to transport
matter through the opacity gap at a rate capable of fueling a bright AGN. 


\section{Appendix B: Disk Stability}
\label{appendix:stability}

Appendix \ref{appendix:equations} lists the full set equations we
solve for calculating the disk models presented in this paper.
Because of the complicated temperature dependence of the opacity
(Fig.~\ref{plot:kp}), together with equations (\ref{hydroapp}) and
(\ref{tempfuncapp}), we find multiple solutions at some radii. These
solutions have the same pressure (by eq. [\ref{hydroapp}]), but
different temperatures and opacities.  In order to determine whether
all of these solutions are physical, we assess the thermal and viscous
stability of our disk solutions in this Appendix.  A second motivation
for doing so is that radiation pressure dominated disks close to black
holes are known to be thermally and viscously unstable in certain
circumstances (e.g., Piran 1978), and so it is important to check
whether the same is true for our solutions.

\subsection{B.1 Timescales}

There are four relevant timescales characterizing the disk: (1) the
dynamical timescale, $\tau_{\rm dyn}\sim\Omega^{-1}$, (2) the
advection timescale, $\tau_{\rm adv}=r/V_r$, (3) the star formation
timescale, $\tau_\star=(\eta\Omega)^{-1}$, and (4) the thermal or
cooling timescale, $\tau_{\rm th}=\Sigma_g c_s^2/F$, where $F$ is the
flux.  The latter is also the photon diffusion time across the disk.

In the optically-thick limit, and assuming that radiation pressure
dominates gas pressure ($p_{\rm rad}/p_{\rm gas}\gg1$), the ratio of
the cooling timescale to the dynamical timescale is very small: \be
\frac{\tau_{\rm th}}{\tau_{\rm
dyn}}\sim\left(\frac{c_s}{c_{}}\right)\tau_{\rm V}.  \ee For the disk
solutions presented in Figure \ref{plot:mds}, the optical depth is
always greater than unity, but not larger than $\sim1000$, even for
$r\sim1$ pc.  Typically, the optical depth is closer to $\sim10-100$.
This, together with the fact that $c_s\sim f_g\sigma$
(eq.~[\ref{cszero}]) implies that $\tau_{\rm th}/\tau_{\rm dyn}\ll1$.
This is an important difference relative to canonical accretion disk
models.  For a local $\alpha$ viscosity an accretion-heated disk has
$\tau_{\rm th} \sim \alpha^{-1} \tau_{\rm dyn} \gg \tau_{\rm dyn}$.

We may also compare the star formation timescale with the dynamical
timescale.  Again assuming that $p_{\rm rad}/p_{\rm gas}\gg1$ and that
$\tau_{\rm V}>1$, we find that \be \frac{\tau_{\star}}{\tau_{\rm
dyn}}\sim\left(\frac{c}{c_s}\right)\epsilon\tau_{\rm V}.  \ee As a
consequence, throughout our disk models $\tau_\star/\tau_{\rm
dyn} > 1$.  This result is, of course, in keeping with our assumption
that star formation occurs on a timescale longer than the local
free-fall timescale (that is, $\eta\leq1$).  

Lastly, we compare the advection timescale with the dynamical
timescale.  If angular momentum transport is produced by a global
torque (as in Figure \ref{plot:mds}), $\tau_{\rm adv} \sim \tau_{\rm
dyn}(r/h) m^{-1}$, which is always greater than unity.  If angular
momentum transport is instead driven by local viscosity, $(\tau_{\rm
adv}/\tau_{\rm dyn})\sim(r/h)^2\alpha^{-1}$.  As discussed in \S3, the
advection timescale can be larger or smaller than the star formation
timescale, depending on the ratio of $\dot M$ to $\dot M_c$.

These considerations lead to the following well-defined hierarchy of
timescales in $Q \sim 1$ disks: \be \tau_{\rm th}\ll\tau_{\rm dyn}\ll
\tau_\star.  \ee

\subsection{B.2 Thermal Stability}
\label{section:thermalstab}

The fact that $\tau_{\rm th}$ is much less than $\tau_{\rm dyn}$ has
important implications for the thermal stability of the disk. In
particular, because of this disparity of timescales, the star
formation rate per unit area $\sds$ is constant on a timescale
$\tau_{\rm th}$.  This implies that the volumetric heating rate
$q^+\sim F/h\propto\sds/h$ is a constant on $\tau_{\rm th}$, and,
hence, $d\ln q^+/d\ln T=0$.  However, the volumetric cooling rate in
the optically thick limit is $q^-\sim \sigma_{\rm SB}T^4/(\tau_{\rm
V}h)$.  Taking $\kappa\propto T^{\beta}$, $d\ln q^+/d\ln T \propto
(4-\beta)$.  Therefore, if $\beta > 4$, the solution is thermally
unstable.  This condition is only obtained when $\kappa\propto T^{10}$
at $T\sim2000$ K (see Figure \ref{plot:kp}; Bell \& Lin 1994).  All
other solutions are thermally stable.  In particular, our $Q \sim 1$
radiation-pressure supported starburst disk with $T \sim 100$s K is
globally thermally stable, in contrast to radiation-pressure supported
disks heated by accretion energy.

In the optically thin limit, the cooling rate is $q^-\sim j\sim\kappa
B\propto T^{4+\beta}$, where $j$ is the emissivity.  Therefore, if
$\beta < -4$, the solution is unstable.  There are only two regions
where such a condition might obtain and they correspond to sudden
decreases in opacity (nearly step functions in $\kappa[T]$) at
$T\sim100-200$ K and in the opacity gap at $T\sim1000$ K in Figure
\ref{plot:kp}.  Otherwise, all optically-thin solutions are globally
thermally stable.

\subsection{B.3 Viscous Stability}
\label{section:viscousstab}

To evaluate the viscous stability of our disks we must compute the
partial derivative $\p\dot{M}/\p\Sigma_g$ (e.g., Lightman \& Eardley
1974).  For $\p\dot{M}/\p\Sigma_g<0$, the disk is viscously unstable
and we expect perturbations to grow and disrupt the disk on a
timescale $\tau_{\rm adv}$.  If angular momentum transport is driven
by global torques, $\dot{M}\propto\Sigma_g c_s$.  But as a consequence
of our assumption of marginal stability against self-gravity
($Q\sim1$), at any radius, $c_s\propto f_g\propto \dot{M}^{1/2}$.
Moreover, because $\tau_{\rm adv} \gg \tau_{\rm dyn}$ it is reasonable
to maintain $Q \sim 1$ during perturbations to the surface density.
These results imply that $\p\dot{M}/\p\Sigma_g$ is always greater than
zero and the disk is viscously stable.  A similar argument applies if
angular momentum transport is instead driven by local viscosity
($\dot{M}\propto\Sigma_g c_s^2$).  $Q \sim 1$ disks thus appear to be
viscously stable.

\subsection{B.4 Physical Disk Solutions} 

In the region where $Q \sim 1$, we often find three solutions to our
disk equations, one low-temperature optically-thick solution and two
high-temperature optically-thin solutions (see also Sirko \& Goodman
2003).  Throughout this paper we have focused on the low-temperature
optically-thick solution as the physical solution.  Here we justify
this choice.

All three of the of the disk solutions appear viscously stable.  The
two high-temperature solutions bracket the opacity gap: one has
$T=T_{\rm sub}\sim1000$ K and the other occurs at $T\sim2000-5000$ K.
By the optically-thin thermal stability criteria of
\S\ref{section:thermalstab}, the solution with $T\sim1000$ K is
thermally unstable because $\beta$ is very large and negative.  This
solution is therefore probably unphysical.  The solution with
$T\sim2000-5000$ K is formally thermally stable because $\beta$ is
positive (Fig.~\ref{plot:kp}).  However, the opacity curve employed
here does not take into account important line cooling processes in
this temperature range.  Indeed, studies of the ISM have shown that
gas with $T\sim2000-5000$ K is thermally unstable (Wolfire et
al.~1995).  Although the ISM of dense starburst
nuclei may be interestingly different than that of normal star-forming
galaxies, it is still likely that optically thin gas at several
thousand Kelvin is thermally unstable.  For this reason we do not
believe that the $T\sim2000-5000$ K solution represents a physical
global equilibrium of the disk.  It is, however, likely that the
existence of multiple solutions at the same pressure implies that the
ISM is prone to breaking up into a multi-phase medium.  A detailed
investigation of the multi-phase structure of the ISM in starburst
galaxies is clearly of interest, but beyond the scope of this paper.

\section{Appendix C: Disk Equations}
\label{appendix:equations}

The equations employed in constructing the models shown in Figures \ref{plot:mds}-\ref{plot:spect}
and described in \S\ref{section:disk} are
\be
\Omega(r)=\Omega_{\rm K}(r)=\left(\frac{GM_{\rm BH}}{r^3}+\frac{2\sigma^2}{r^2}\right)^{1/2}
\ee
\be
\sds=\Sigma_g\Omega\eta
\ee
\be
p_{\rm gas}+\epsilon\sds c\left(\frac{1}{2}\tau_{\rm V}+\xi\right)=\rho h^2\Omega^2
\label{hydroapp}
\ee
\be
p_{\rm gas}=\rho k_{\rm B}T/m_p
\ee
\be
T^4=\frac{3}{4}\teff^4\left(\tau_{\rm V}+\frac{2}{3\tau_{\rm V}}+\frac{4}{3}\right)
\label{tempfuncapp}
\ee
\be
\tau_{\rm V}=\kappa\Sigma_g/2
\ee
\be
\Sigma_g=2\rho h
\ee
\be
\dot{M}=4\pi Rh\rho V_r=4\pi Rh\rho mc_s=4\pi Rh^2\rho\Omega m
\label{dynapp}
\ee \be \dot{M}=\dot{M}_{\rm out}-\int_{R_{\rm out}}^r 2\pi r \sds dr
\ee In the outer part of the disk where accretion heating is
insufficient to maintain $Q \sim 1$, these equations are solved
subject to the conditions \be \rho=\frac{\Omega^2}{\sqrt{2}\pi GQ}
\label{qapp}
\ee
(with $Q=1$) and
\be
\sigma_{\rm SB}\teff^4=\frac{1}{2}\epsilon\sds c^2+
\frac{3}{8\pi}\dot{M}\left(1-\sqrt{R_{\rm in}/r}\right)\Omega^2.
\label{teffapp}
\ee In the inner part of the disk where accretion heating maintains $Q
> 1$, equations (\ref{qapp}) and (\ref{teffapp}) are replaced by \be
\sigma_{\rm SB}\teff^4=\frac{3}{8\pi}\dot{M}\left(1-\sqrt{R_{\rm in}/r}\right)\Omega^2.  \ee The radius
at which $Q > 1$ is denoted $R_{\rm AGN}$.  Throughout this work we
set $\xi=1$ in equation (\ref{hydroapp}) and $\epsilon=10^{-3}$ in
equation (\ref{teffapp}).  $M_{\rm BH}$ is specified by $M_{\rm
BH}=2\times10^8 \sigma_{200}^4$ M$_\odot$.  The inner edge of the AGN
disk is taken to be at $R_{\rm in}=3(2GM_{\rm BH})/c^2$.
$\dot{M}_{\rm out}$ and $R_{\rm out}$ are the mass supply rate at the
outer boundary and the location of the outer boundary, respectively.
In equation (\ref{dynapp}), we generally take $m=0.1-0.2$.

When calculating disks whose accretion is driven by a local viscosity, as in Figure \ref{plot:gc},
we replace equation (\ref{dynapp}) with
\be
\dot M = 2 \pi \nu \Sigma_g \left|\frac{d\ln\Omega}{d\ln r}\right|=\frac{2^{3/2}\alpha h^3 \Omega^3}{GQ}\left|\frac{d\ln\Omega}{d\ln r}\right|.
\label{mdotviscapp}
\ee 

\section{Appendix D: The Critical Mass Supply Rate}
\label{appendix:mdotcrit}

\begin{figure}
\begin{center}
\plotone{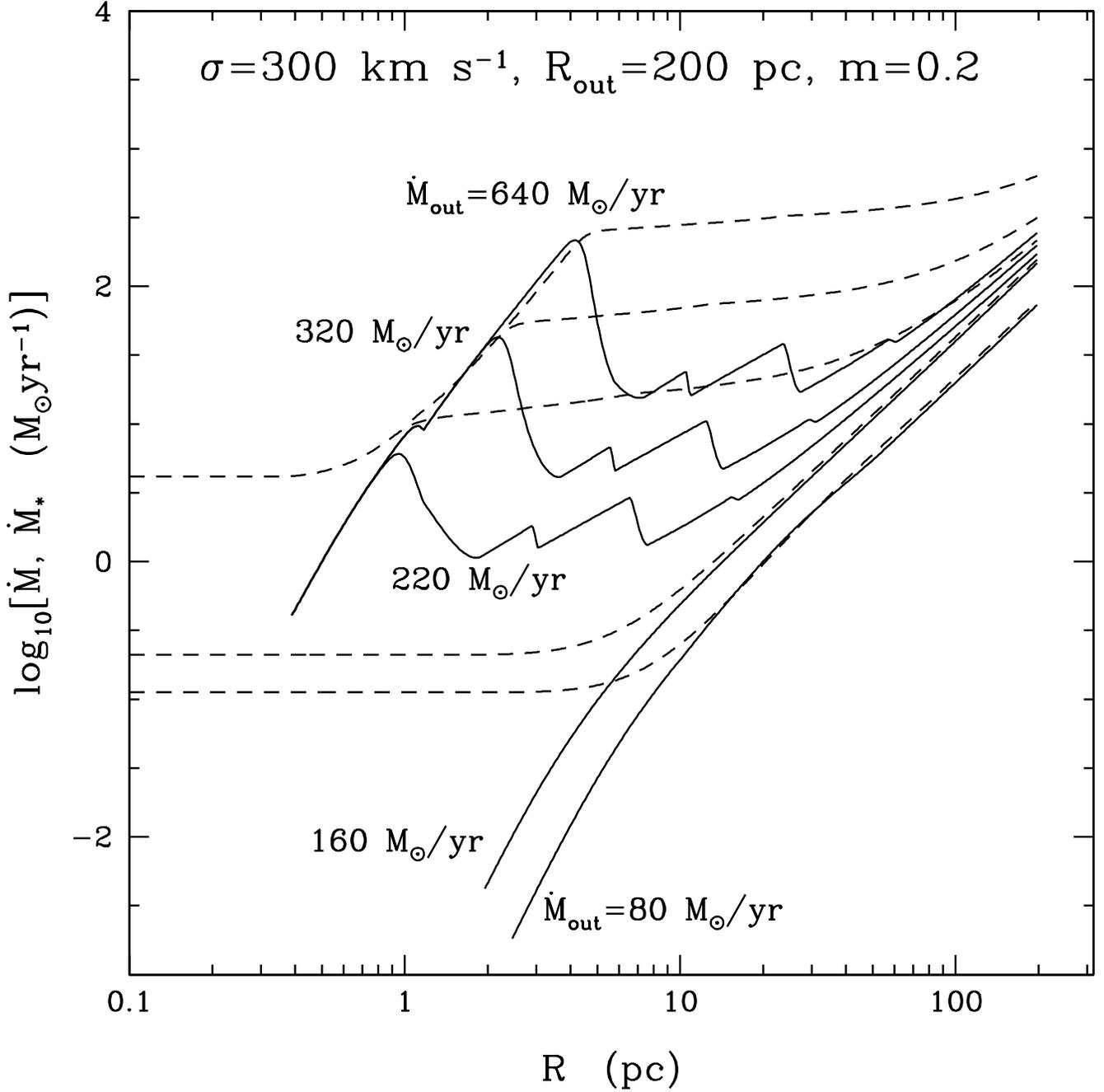}
\end{center}
\figcaption{The local star formation rate ($\mds=\pi r^2 \sds$; solid
lines) and accretion rate (dashed lines) as a function of radius for
$\dot{M}_{\rm out}=80$, 160, 220, 320, and 640 M$_\odot$ yr$^{-1}$ in
a model with $R_{\rm out}=200$ pc and $\sigma=300$ km s$^{-1}$.  Note
the strong bifurcation between models with low mass supply rates
($\dot{M}_{\rm out}\le160$ M$_\odot$ yr$^{-1}$; see eq.~[\ref{mdotc}])
and models with high mass supply rates ($\dot{M}_{\rm out}\ge220$
M$_\odot$ yr$^{-1}$).  The latter produce outer ($r\sim R_{\rm out}$)
and inner ($r\sim1-10$ pc) starbursts, but also fuel a bright central
AGN with $\dot{M}_{\rm BH}\sim4$ M$_\odot$ yr$^{-1}$ (Appendix
\ref{appendix:accrete}).  Figure \ref{plot:spect} shows the computed
spectra for each model.  Figure \ref{plot:allp} shows the full disk
structure for the model with $\dot{M}_{\rm out}= 320 $M$_\odot$
yr$^{-1}$.
\label{plot:mds}}
\end{figure}

\begin{figure}
\begin{center}
\plotone{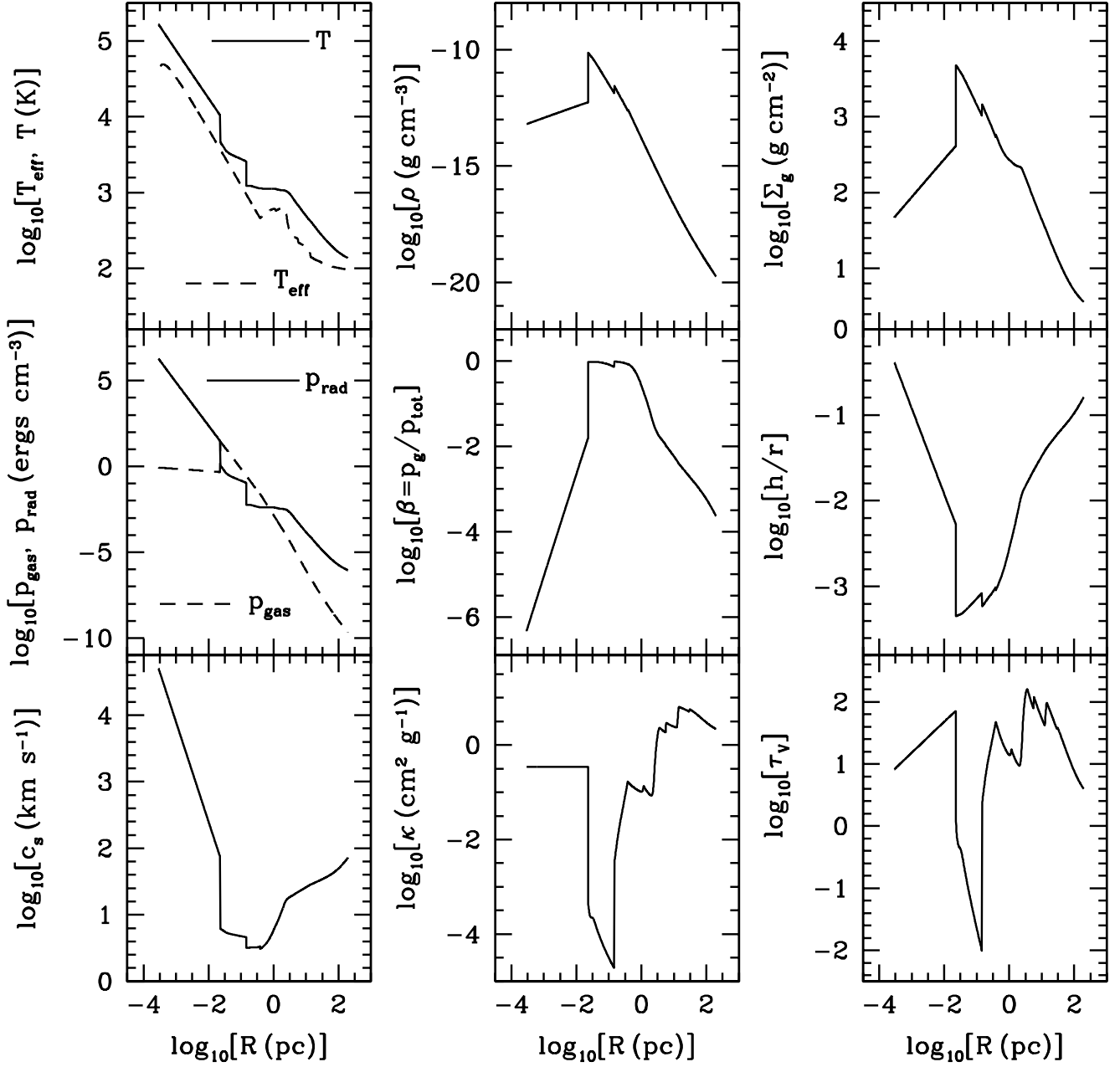}
\end{center}
\figcaption{$\log_{10}$ of $T$ (solid line, upper left panel), $\teff$
(dashed line, upper left panel), $\rho$ (upper middle panel),
$\Sigma_g$ (upper right panel), radiation pressure $p_r$ (solid line,
middle left panel), gas pressure $p_g$ (dashed line, middle left
panel), $\beta=p_{\rm g}/p_{\rm tot}$ (middle panel), $h/r$ (middle
right panel), $c_s$ (lower left panel), $\kappa$ (lower middle panel),
and $\tau_{\rm V}$ (lower right panel) all in cgs units as a function
of radius in the model with $\dot{M}_{\rm out}=320$ M$_\odot$
yr$^{-1}$, $R_{\rm out}=200$ pc, and $m=0.2$.
The opacity gap region, highlighted in Figure
\ref{plot:mds}, is located at $r\sim1$ pc.  
\label{plot:allp}}
\end{figure}

\begin{figure}
\begin{center}
\plotone{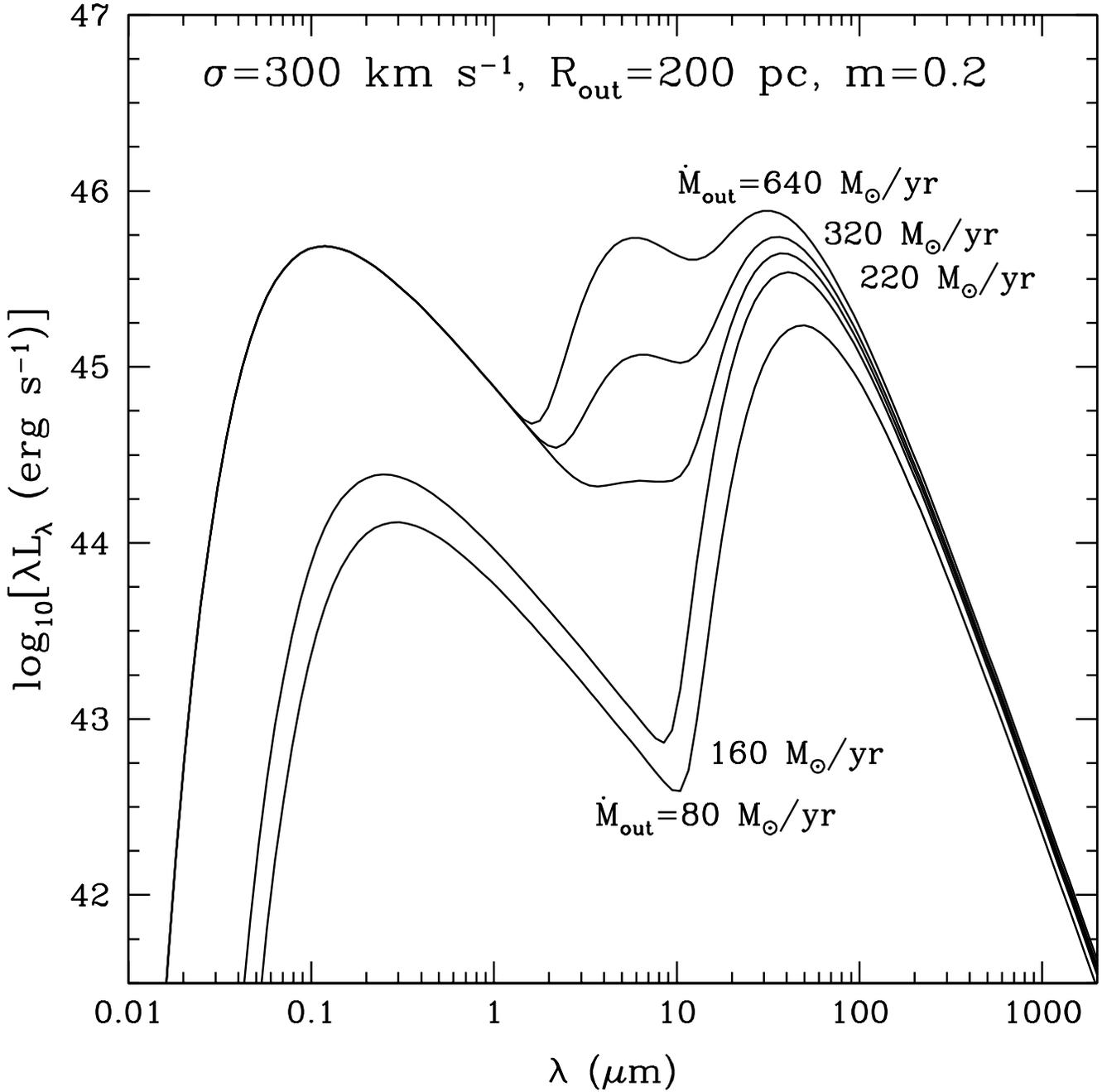}
\end{center}
\figcaption{The spectral energy distribution as a function of wavelength
for each of the disk solutions presented in Figure \ref{plot:mds}.
Solutions with small mass supply rates ($\dot{M}_{\rm out}\le160$
M$_\odot$ yr$^{-1}<\dot{M}_c$) are dominated by a starburst at large
radii that produces a prominent FIR peak.  Solutions with large mass
supply rates ($\dot{M}_{\rm out}\ge220$ M$_\odot$ yr$^{-1}>\dot{M}_c$)
have a similar FIR bump, but also have a spectral peak at $\sim10$
$\mu$m due to the inner ring of star formation in the opacity gap at
$1-10$ pc.  As shown in Figure \ref{plot:mds}, these solutions also
provide mass to the black hole at a rate comparable to the Eddington
limit.  The total bolometric emission from these models can be
dominated by the central AGN, although there is a significant
contribution from star formation (Appendix \ref{appendix:accrete}).
\label{plot:spect}}
\end{figure}

\begin{figure}
\begin{center}
\plotone{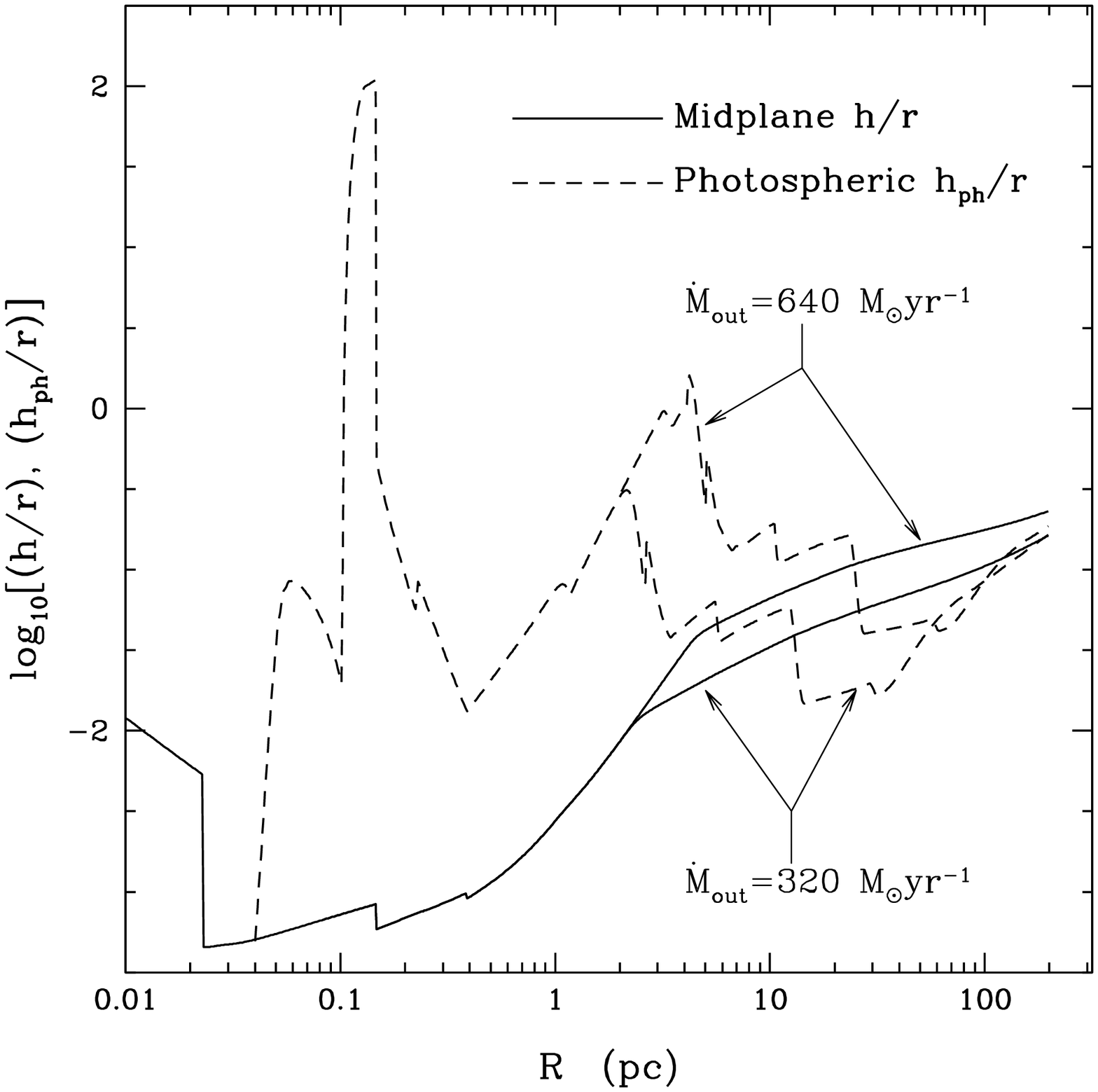}
\end{center}
\figcaption{Estimates of the midplane (solid lines) and photospheric
(dashed lines) scale-heights for two models with $\dot{M}_{\rm
out}=320$ and 640 M$_\odot$ yr$^{-1}$.  The model with $\dot{M}_{\rm
out}=640$ M$_\odot$ yr$^{-1}$ has higher midplane $h/r$ at $R_{\rm
out}$ and higher photospheric $h_{ph}/r$ at $r\sim10$ pc.  Both models
have the same midplane and photospheric scale height for $r\lesssim2$
pc because of the power law solution in the opacity gap (Appendix
\ref{appendix:accrete}).  Note that for $0.1\lesssim r \lesssim0.5$ pc
the disk is gas pressure dominated at the midplane, but radiation
pressure dominated at the photosphere (see Fig.~\ref{plot:allp}).
Values of $h_{ph}/r \gtrsim 1$ indicate that the photosphere is
unbound. \label{plot:hr}}
\end{figure}

\begin{figure}
\begin{center}
\plotone{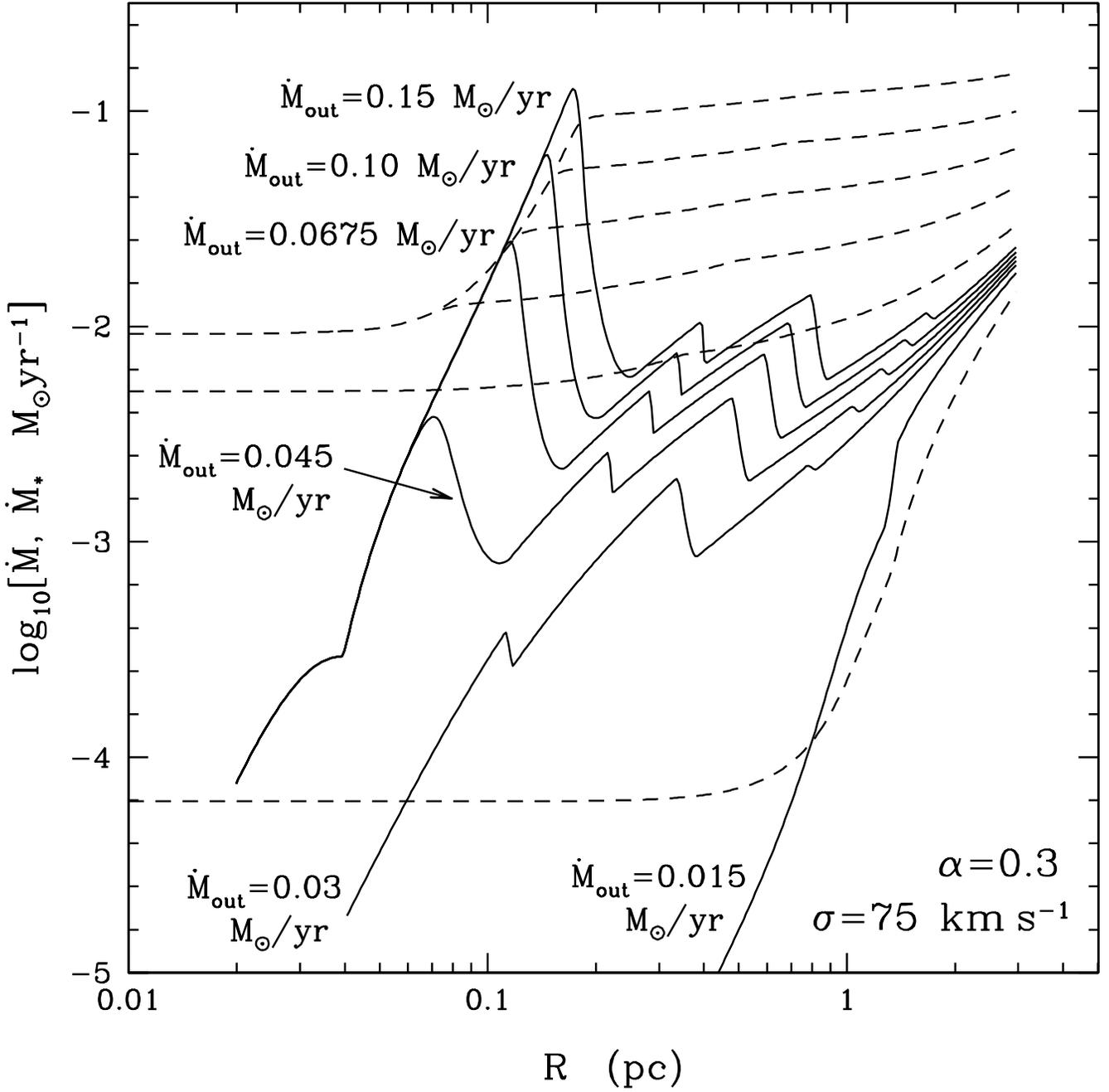}
\end{center}
\figcaption{Star formation rate (solid lines) and gas accretion rate (dashed lines) in disk models
appropriate to the Galactic Center ($M_{\rm BH} = 4 \times 10^6 M_\odot$
and $\sigma = 75 \kms$).  We consider a mass supply rate $\dot M_{\rm
out}$ at a radius $R_{\rm out} = 3$ pc and accretion via a local
$\alpha$ viscosity with $\alpha = 0.3$.  The chosen values of $R_{\rm
out}$ and $\dot M_{\rm out}$ are motivated by observations of the
circumnuclear disk in the Galactic Center.  Note that the opacity was increased
by a factor of 3 relative to Figure \ref{plot:kp} to account for the
super-solar metallicity in the Galactic Center.  These results
show a ring of star formation at $\sim 0.1$ pc where there is
currently a young disk of O and B stars.\label{plot:gc}}
\end{figure}

As shown in Figure \ref{plot:mds} and Appendix \ref{appendix:accrete},
when $\dot{M}_{\rm out}$ exceeds a critical rate $\dot{M}_c$, gas can
accrete to small radii fueling a bright AGN. By contrast, for
$\dot{M}_{\rm out}<\dot{M}_c$, star formation consumes the gas in a
narrow range of radii at $\sim R_{\rm out}$.  In the optically-thick
limit, and to the extent that $\kappa$ may be approximated by
$\kappa_0T^2$, the critical mass supply rate is given in equation
(\ref{mdotc}).

The balance between star formation and accretion is different in the
optically-thin limit than it is in the optically-thick limit.
To show this, we note that equation (\ref{mdotevol}) admits a simple
power-law solution when angular momentum transport is driven by a
global torque (eq.~[\ref{mdotdyn}]):\be \dot{M}(r)\propto
r^{\sigma/(\sqrt{2}\epsilon\xi c m)}.
\label{mdotcthin}
\ee Because \be \frac{\sigma}{\sqrt{2}\epsilon\xi c
m}\sim7\sigma_{300}(\epsilon_3\xi m_{0.1})^{-1} \ee we see that unless
there is a very efficient global torque ($m \gtrsim 1$), gas will be
converted into stars rather than accreting to smaller radii.

\end{appendix}


\end{document}